\documentclass[11pt]{article}

\usepackage[T1]{fontenc}
\usepackage[utf8]{inputenc}
\usepackage[a4paper,margin=1in]{geometry}

\usepackage{setspace}
\usepackage{authblk}

\usepackage{graphicx}
\usepackage{caption}
\usepackage{float}
\graphicspath{{figures/}{fig/}{Si Fig/}{./}}

\usepackage{amsmath,amssymb}
\usepackage[version=3]{mhchem}
\usepackage{textgreek}
\usepackage{siunitx}
\usepackage{achemso}
\setkeys{acs}{doi=true}

\usepackage[hidelinks]{hyperref}

\title{Lipid Hydrocarbon Tail Structure Governs Interfacial Anchoring and Stripe Morphology in Cholesteric Liquid Crystals}

\author[1]{Mengwei Li}
\author[1,2]{Stefanie D. Pritzl}
\author[3]{Martin F. Haase}
\author[1]{Lisa Tran\thanks{Corresponding author: \texttt{l.tran@uu.nl}}}

\affil[1]{Soft Condensed Matter and Biophysics, Debye Institute for Nanomaterials Science, Utrecht University, 3584 CC Utrecht, The Netherlands}
\affil[2]{Institute of Physics, University of Augsburg, Universit\"{a}tsstrasse 1, 86179 Augsburg, Germany}
\affil[3]{Van't Hoff Laboratory of Physical and Colloid Chemistry, Debye Institute for Nanomaterials Science, Utrecht University, 3584 CH Utrecht, The Netherlands}

\date{}

\begin{document}
\maketitle

\begin{abstract}
\noindent Liquid crystal-based biosensors rely on the exceptional sensitivity of interfacial anchoring to molecular adsorption, and cholesteric liquid crystals are particularly powerful in this context because their helical structure supports multiple optically distinct textures, such as fingerprint stripes, distorted helices, and uniform homeotropic states, that vary continuously with anchoring strength. Despite the widespread use of amphiphiles to control liquid crystal alignment, the molecular origins by which lipid acyl chain structure regulates cholesteric anchoring and pattern selection remain unclear. Here, we systematically compare saturated 1,2-dilauroyl-sn-glycero-3-phosphocholine (DLPC) and unsaturated 1,2-dioleoyl-sn-glycero-3-phosphocholine (DOPC) monolayers to determine how hydrocarbon tail composition governs interfacial organization and director alignment. By mapping the interfacial stripe spacing across lipid concentration, mixing ratio, and cholesteric pitch, we find that the transition from periodically modulated fingerprint textures to uniform homeotropic alignment is consistent with differences in lipid packing efficiency and lateral organization at the interface. The behavior of saturated DLPC is consistent with the formation of relatively dense and homogeneous interfacial layers that efficiently transmit orientational constraints and promote unwinding of the cholesteric helix. In contrast, the behavior of DOPC suggests that \textit{cis}-double bonds introduce steric incompatibility that delays this transition and generates spatial heterogeneity in stripe morphology. Fluorescence Recovery After Photobleaching (FRAP) measurements provide complementary information on interfacial fluorescent-probe mobility. Probe mobility becomes strongly restricted at high lipid concentration for both systems, consistent with increasing lateral constraint as surface coverage increases. DOPC-containing interfaces exhibit higher diffusion at low coverage due to greater chain flexibility, while increasing surface density suppresses diffusion for both lipids, reflecting the formation of densely packed layers that enhance collective orientational coupling. Variations in cholesteric pitch and film thickness further modulate these effects, leading to coexisting textures and, under certain confinement conditions, features consistent with lateral lipid heterogeneity at the interface. Together, these results connect lipid acyl chain structure and its influence on interfacial packing and mobility to the anchoring of cholesteric liquid crystals, providing molecular-level design principles for responsive liquid crystal interfaces.
\end{abstract}


\newpage
\section{Introduction}

\indent \indent Liquid crystal-based biosensors are label-free optical sensors that transduce molecular adsorption at a fluid interface into a macroscopic change in liquid crystal orientation, allowing species ranging from surfactants and lipids to proteins and nucleic acids to be detected without molecular labels or complex instrumentation \cite{carlton2013chemical, qu2022overview, gupta1997optical, Brake2003BiomolecularInteractions}. The orientational state of a liquid crystal is described by the \textit{director}, a unit vector representing the local average orientation of its elongated molecules, while the spatial variation of this orientation defines the \textit{director field}. Molecular adsorption alters this director field through interfacial anchoring, the boundary condition that sets the preferred director orientation at a surface \cite{jerome1991surface, de1993physics}. When amphiphilic species such as surfactants, phospholipids, or proteins assemble at a liquid crystal-aqueous boundary, they modify the anchoring energy and reorient the interfacial director. Because the liquid crystal is birefringent, this reorientation changes how the interface transmits polarized light, amplifying a molecular-scale adsorption event into a macroscopic change in brightness, color, or texture visible under a polarizing microscope \cite{gupta1997optical, brake2002experimental, brake2003active, meli2008preparation}. In many systems, increasing surface adsorption drives the director from planar (parallel) to homeotropic (perpendicular) alignment relative to the interface \cite{lockwood2005influence, moreno2012liquid}. This coupling between molecular-scale adsorption and bulk optical response forms the foundation of liquid crystal-based detection strategies.

Beyond serving as optical sensors, liquid crystals can be used to probe interfacial molecular organization that is otherwise difficult to access directly. Molecular-scale changes in amphiphile adsorption, packing, lateral organization, or local composition alter the anchoring condition at the liquid crystal-aqueous interface, and the liquid crystal converts these changes into mesoscale director configurations and optical textures that are directly and quantitatively imaged by microscopy. The texture is thus a direct readout of the interfacial anchoring, whereas the molecular-scale origin of that anchoring, the specific lipid packing, local composition, and chain conformation, is inferred from the texture. It is in this sense that we describe the liquid crystal as an indirect reporter of interfacial molecular interactions.

While nematic liquid crystals have been widely studied for molecular sensing applications, cholesteric liquid crystals offer additional functionality due to their intrinsic helical structure \cite{bisoyi2021liquid, de1993physics}. In cholesterics, the competition between bulk twist elasticity and surface anchoring generates a rich landscape of interfacial textures, including fingerprint stripes and other periodically modulated states, whose spacing and morphology depend sensitively on anchoring strength and geometric confinement \cite{sec2012geometrical, guo2016cholesteric, poy2017role, tran2017change}. Because these textures evolve continuously between planar-modulated and homeotropic states, cholesterics provide multiple optically distinct signatures within a single material system. This multiplicity of accessible patterns enhances their potential as responsive platforms and enables more nuanced readouts of interfacial events \cite{tran2018shaping, pawale2025directed, ren2025stretchable}.

The richer set of cholesteric textures provides an opportunity to examine whether interfacial anchoring is determined solely by the amount of adsorbed amphiphile or also by its molecular structure and lateral organization. We investigate this question using phospholipids as model amphiphiles, whose biological relevance and systematically tunable acyl chain chemistry make them well suited for examining how molecular structure regulates interfacial packing \cite{van2008membrane,harayama2018understanding}. Lipids vary in acyl chain length and degree of saturation, and these molecular features strongly influence monolayer packing, mobility, and lateral organization \cite{israelachvili2011intermolecular,nagle2000structure}. Saturated lipids such as 1,2-dilauroyl-sn-glycero-3-phosphocholine (DLPC) adopt relatively extended hydrocarbon-chain conformations that favor efficient packing, whereas unsaturated lipids such as 1,2-dioleoyl-sn-glycero-3-phosphocholine (DOPC) contain \textit{cis}-double bonds that introduce persistent chain kinks and increase conformational disorder. Although DOPC has longer C18 chains than the C12 chains of DLPC, studies of fluid phosphatidylcholine membranes show that \textit{cis}-unsaturation can substantially increase chain flexibility and effective molecular area and thereby counteract the ordering expected from increased chain length \cite{rawicz2000effect,kusumi1986spin,binder2001effect}. Thus, while the present comparison does not independently separate chain length from saturation, the different interfacial organization of DLPC and DOPC is expected to be strongly influenced by their difference in acyl-chain saturation. These structural differences may cause lipid layers with comparable surface coverage to impose anchoring conditions that differ in strength and spatial uniformity. Although prior work on nematic interfaces has shown that hydrocarbon chain structure can influence anchoring transitions \cite{brake2003effect,lockwood2005influence}, how acyl-chain geometry and packing couple specifically to the twist elasticity of cholesteric liquid crystals remains largely unexplored.

Here, we address this question by systematically comparing saturated and unsaturated phospholipids at cholesteric-aqueous interfaces to determine how hydrocarbon tail structure influences interfacial organization and anchoring behavior. By mapping the characteristic interfacial stripe spacing across lipid concentration, mixing ratio, and cholesteric pitch, we examine how lipid structure modifies the evolution from periodically modulated fingerprint textures toward predominantly homeotropic alignment. In parallel, Fluorescence Recovery After Photobleaching (FRAP) measurements \cite{loren2015fluorescence,day2012analysis} probe the lateral mobility of a fluorescent phospholipid tracer at the interface. Through this combined structural and dynamical analysis, we show that lipid acyl chain structure is associated with differences in anchoring strength, texture regularity, and interfacial mobility.

\newpage

\section{Results and Discussion}
\subsection{Effect of lipid hydrocarbon tail geometry in anchoring cholesterics}
\indent \indent To compare lipid-induced anchoring in a cholesteric system, we used 5CB (4-cyano-4$'$-pentylbiphenyl) doped with 2.8 wt.-\% of the chiral additive CB15 ((S)-4-cyano-4$'$-(2-methylbutyl) biphenyl), producing a cholesteric phase with a helical pitch of approximately 5~µm \cite{bisoyi2021liquid,tran2017change,tran2018shaping,tran2020swelling}. The pitch, $p$, is the distance over which the liquid crystal director completes a full $360^\circ$ rotation. When the cholesteric helix lies parallel to the aqueous interface, this rotation produces the characteristic fingerprint texture, with alternating director orientations repeating approximately every half-pitch ($p/2\approx2.5$~µm) in the absence of strong distortion. Lipid-induced homeotropic anchoring competes with this intrinsic twist. Increasing anchoring progressively distorts and locally unwinds the helix near the interface, altering the stripe morphology and spacing before ultimately favoring a predominantly homeotropic state. This relationship between the intrinsic pitch and lipid-induced helix distortion provides the basis for using the cholesteric texture as a sensitive reporter of interfacial anchoring. Saturated DLPC (C12:0) and unsaturated DOPC (C18:1) were selected as representative amphiphiles with distinct hydrocarbon tail geometries (Fig.~S1A) \cite{israelachvili2011intermolecular,nagle2000structure,pinot2014polyunsaturated}. Because DLPC and DOPC differ in both chain length and degree of saturation, this comparison does not isolate the effect of either variable independently. We therefore also examine a series of mixed DLPC:DOPC compositions to investigate how their relative contributions influence interfacial anchoring. Texas Red-DHPE (TR-DHPE; Fig.~S1A) was incorporated at a concentration of 1 mol-\% to visualize interfacial lipid organization by confocal fluorescence microscopy.

To stabilize the liquid crystal interface for imaging and ensure well-defined confinement, we adopted an experimental geometry based on a transmission electron microscopy (TEM) grid supported on an octadecyltrichlorosilane (OTS)-treated glass slide (Fig.~S1B). This configuration builds directly on methodologies developed by Brake and co-workers, which enable controlled observation of lipid-liquid crystal interactions within microscale grid compartments \cite{brake2003active,tran2018shaping}. In this geometry, lipids assemble at the aqueous interface with their hydrophilic headgroups exposed to water and hydrophobic tails intercalated into the liquid crystal phase, thereby imposing homeotropic anchoring \cite{tran2017change,lin2011endotoxin,meli2008preparation,lockwood2005influence}. 

When a cholesteric liquid crystal is confined by a surface that promotes homeotropic alignment, its intrinsic helical twist competes with the imposed boundary condition, generating elastic frustration. Previous studies have shown that this competition can produce a progression from planar alignment through periodically modulated fingerprint textures to predominantly homeotropic states as the effective homeotropic anchoring strength increases \cite{tran2017change,tran2018shaping}. Within this established framework, weak anchoring allows the cholesteric to remain predominantly planar at the interface, whereas intermediate anchoring produces alternating planar and homeotropic regions as the director retains part of its preferred helical rotation. When the helical axis lies in the plane of the interface, the surface director cycles between perpendicular and in-plane orientations over approximately half of the intrinsic pitch, giving rise to the characteristic fingerprint periodicity. At sufficiently strong homeotropic anchoring, the helix becomes increasingly distorted and displaced from the interface, producing a predominantly homeotropic state in which the remaining twist is concentrated near isolated defect lines. We use this framework below to interpret the concentration-dependent textures produced by DLPC and DOPC.

This framework predicts how increasing surface coverage progressively perturbs the intrinsic cholesteric twist, driving the progression from planar alignment to fingerprint textures with modified stripe spacing and, ultimately, to a predominantly homeotropic state as the helix is displaced from the interface. We next examine whether that progression depends on lipid molecular structure by comparing DLPC and DOPC across a range of concentrations. Following established liquid crystal sensing protocols \cite{price2008dna,lin2011endotoxin}, we track how each lipid modifies the concentration-dependent evolution of the cholesteric texture (Fig. 1), with particular attention to the onset of fingerprint ordering, the spatial regularity of the stripes, and the approach toward helix unwinding. To minimize variations in local confinement, all representative micrographs in Fig. 1 were acquired from grid squares near the center of the TEM grid.

For consistency in describing these textures, we use \textit{fingerprint texture} to refer to the periodic stripe pattern formed when the cholesteric helix lies with its axis parallel to the interface. \textit{Regular} fingerprint textures contain smoothly curved, uniformly spaced stripes, whereas \textit{distorted} fingerprint textures contain locally bent, undulating, or irregularly spaced stripes associated with spatial variations in effective anchoring. We use \textit{lipid-enriched regions} to describe broadened homeotropically anchored domains in which fluorescent lipids preferentially accumulate.

\begin{figure}[H]
    \centering
    
    \captionsetup{font=scriptsize} 
    \includegraphics[width=1\linewidth]{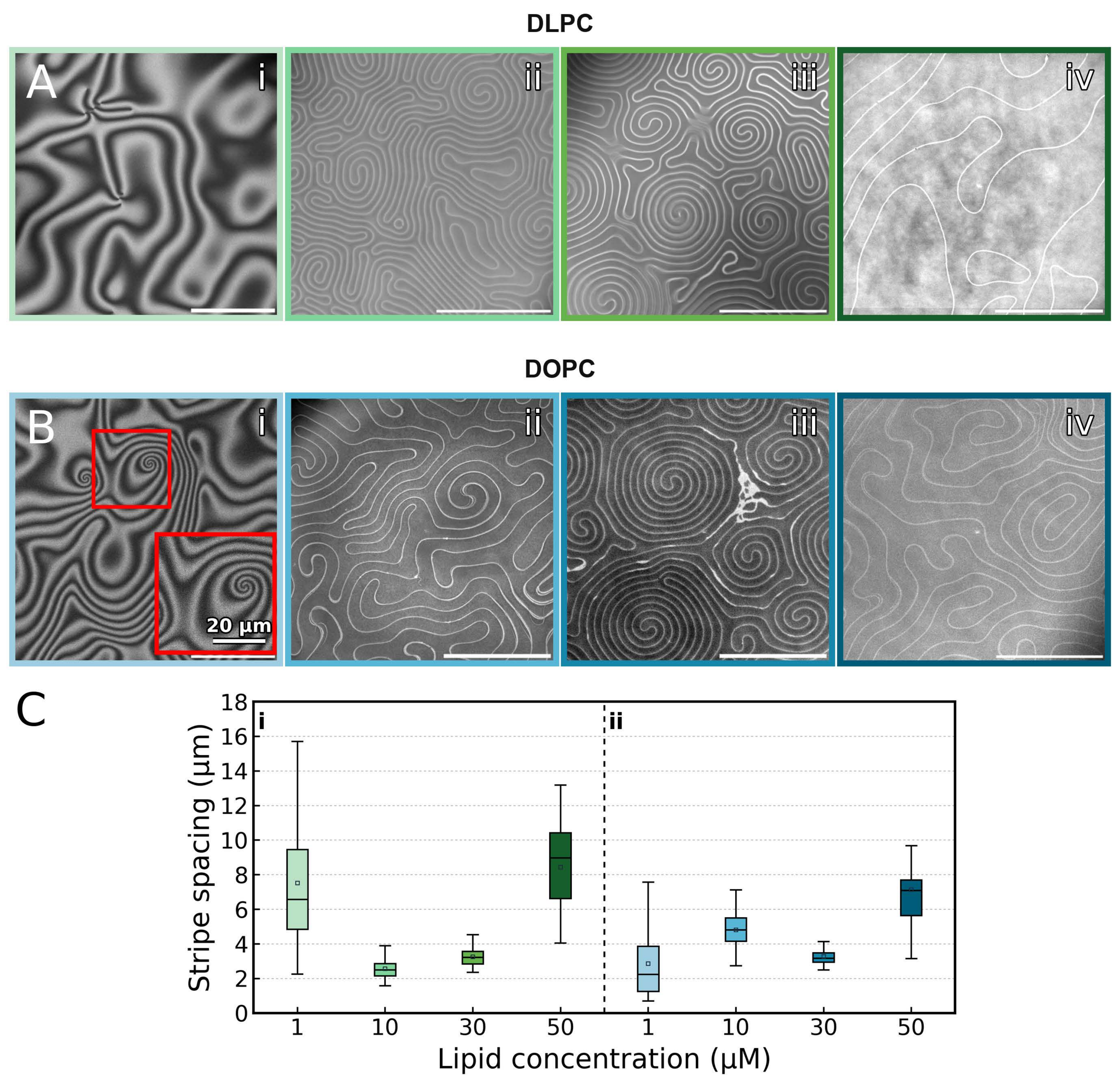}
    \caption{Concentration-dependent interfacial textures and stripe periodicity at the cholesteric liquid crystal-water interface.
(A and B) Representative fluorescence confocal microscopy images of Texas Red-DHPE emission (imaged through the liquid crystal, see Supporting Information) showing the evolution of interfacial textures with varying concentrations: i) 1, ii) 10, iii) 30,  iv) 50 µM of aqueous vesicle suspensions of (A) DLPC (green borders) and (B) DOPC (blue borders) lipids.
(C) Quantitative analysis of the characteristic stripe spacing plotted against lipid concentration. Plot (C-i) corresponds to DLPC (green), and plot (C-ii) corresponds to DOPC (blue). (Scale bar:
50 µm.) A cross-sectional schematic mapping these alternating planar and homeotropic regions in the fingerprint regime and their evolution toward narrow twist-disclination lines as homeotropic anchoring increases (planar versus homeotropic, with the helix axis parallel to the interface) is given in Fig.~S2.}
    \label{fig:1}
\end{figure}

Fluorescence confocal micrographs in Fig.~1A and 1B show the progression from predominantly planar alignment, through fingerprint textures with alternating planar and homeotropic regions, to predominantly homeotropic interfaces separated by narrow twist-disclination lines. Complete image series and corresponding grayscale intensity profiles are provided in the Supporting Information for DLPC and DOPC as functions of lipid concentration (Figs.~S3 and S4) and CB15 concentration (Figs.~S5 and S6). Consistent with previous studies of lipid-liquid crystal interfaces \cite{moreno2012liquid}, the Texas Red-DHPE fluorescence, interpreted together with the water-side imaging controls discussed in the Supporting Information, supports preferential lipid localization within homeotropically anchored regions.

As a caveat before discussing Fig. 1, we point out that the fluorescence intensity in the confocal micrographs should not be interpreted as a quantitative measure of local lipid concentration. As detailed in the Supporting Information, water-side confocal and wide-field fluorescence imaging shows that narrow twist-disclination regions contain less fluorescent lipid than the surrounding homeotropic regions, despite appearing bright when imaged through the cholesteric liquid crystal \cite{tran2018shaping}. This contrast inversion is attributed to optical propagation through the birefringent cholesteric phase. A two-color control using two spectrally distinct DHPE-based fluorescent probes reproduced the same stripe pattern in both fluorescence channels (Fig.~S7), indicating that the observed pattern is not specific to the photophysical behavior of a single fluorophore. The fluorescence images are therefore used primarily to identify the morphology and evolution of the cholesteric textures, while lipid localization is interpreted from the combined confocal and water-side imaging controls. At the same time, the strong dependence of fluorescence contrast on propagation through the cholesteric demonstrates an additional form of optical sensitivity. The anisotropic director field couples to the polarization-dependent excitation and detection of the interfacial fluorophores, converting local orientational structure into measurable fluorescence contrast \cite{smalyukh2001three,lavrentovich2003fluorescence}. Thus, this optical coupling provides an additional readout of interfacial orientational organization.

As shown in Fig.~1A and 1B, the interfacial texture depends strongly on lipid concentration, consistent with previous studies \cite{brake2002experimental,brake2003active,lockwood2005influence}. In pure water, the cholesteric favors planar anchoring at the aqueous interface. At 1~µM, both DLPC and DOPC produce broad, schlieren-like fluorescence textures (Fig.~1A and 1B, column i). These patterns are not visible during water-side imaging, suggesting that they do not primarily represent large-scale variations in lipid concentration. Instead, they likely arise from weak director distortions that modify the propagation, polarization, and collection of fluorescence through the birefringent liquid crystal. At this low surface coverage, lipid adsorption begins to perturb the planar state but remains insufficient to establish a fully developed fingerprint texture with the helical axis parallel to the interface \cite{moreno2012liquid,tran2017change,tran2018shaping,lavrentovich2020undulation,blanc2023helfrich}. The possible optical origin of this schlieren-like contrast is discussed further in the Supporting Information.

At 10 and 30~µM (Fig.~1, columns ii-iii), well-developed fingerprint textures emerge from the competition between intrinsic twist elasticity and lipid-induced homeotropic anchoring, with alternating planar and homeotropic regions spaced according to the cholesteric half-pitch \cite{de1993physics,oswald2005nematic}. Fluorescent lipids preferentially occupy the homeotropic regions, and increasing adsorption progressively widens these regions while narrowing the intervening planar stripes. At higher coverage, the residual planar regions contract into twist-disclination lines that accommodate the twist elastic frustration. Accordingly, at 50~µM (Fig.~1, column iv), the interface is predominantly homeotropic, with only isolated twist-disclination lines remaining \cite{tran2018shaping,tran2017change,brake2002experimental}.

The box plots in Fig. 1C quantify the characteristic stripe spacing as a function of the lipid concentration, measured in ImageJ from distances between pixel intensity peaks along lines perpendicular to the stripes. At 1 µM DLPC (Fig. 1C-i), the relatively large median spacing and broad distribution are consistent with a weakly perturbed, predominantly planar interface. At 10 µM, the spacing reaches its minimum and the distribution narrows, indicating the formation of a regular fingerprint texture with a periodicity approaching the cholesteric half-pitch. The spacing and distribution width increase at 30~µM and further at 50 µM as the homeotropic regions broaden and the intervening planar regions narrow into twist disclination lines. This nonmonotonic trend is consistent with increasing lipid adsorption shifting the balance from bulk twist elasticity toward homeotropic anchoring. The proposed expansion of the lipid-enriched homeotropic regions is inferred from the observed textures and complementary fluorescence controls (see Supporting Information). 

Increasing DOPC concentration produces the same general progression of cholesteric textures as DLPC, but with distinct morphologies and transition concentrations. At 1~µM, DOPC produces broad planar regions together with localized, tightly spaced stripes and small spiral features not observed for DLPC (Fig.~1B-i, inset). This spatial heterogeneity may reflect stronger local homeotropic anchoring from the longer C18 tails where favorable molecular organization occurs, consistent with studies showing that liquid crystal anchoring depends on amphiphile chain length \cite{brake2003effect,lockwood2005influence}.

With increasing DOPC concentration, fingerprint textures become more prominent (Fig.~1B, columns ii-iv), but their spacing remains larger than the expected cholesteric half-pitch ($p/2\sim2.5$~µm for 2.8~wt.-\% CB15 in 5CB), indicating partial helix unwinding (Fig.~1C-ii). Unlike DLPC, however, DOPC does not produce a predominantly homeotropic state with fully unwound twist-disclination spirals at 50~µM. Instead, relatively regular stripes persist, showing that the  cholesteric twist is less frustrated. This behavior is consistent with the longer DOPC tails promoting strong local homeotropic alignment, while their steric bulk and \textit{cis}-double-bond kinks limit the laterally coordinated packing needed to suppress the helix across the interface \cite{israelachvili2011intermolecular}.

The contrasting responses of DLPC and DOPC therefore suggest a concentration-dependent balance between tail-mediated local anchoring and packing-mediated collective anchoring \cite{van2008membrane,harayama2018understanding}. At low coverage, the longer DOPC tails appear to induce homeotropic alignment more effectively \cite{szule2002effects,brake2003effect,lockwood2005influence,skaife1999influence}. At higher coverage, however, the kinked DOPC chains increase molecular cross-sectional area and steric hindrance, limiting dense lateral packing \cite{pinot2014polyunsaturated,israelachvili2011intermolecular,nagle2000structure}. The shorter, saturated DLPC tails instead pack more efficiently as coverage increases, producing weaker anchoring at low concentration but stronger collective homeotropic anchoring at high concentration. Thus, DOPC has more effective homeotropic anchoring locally at low coverage, whereas DLPC more effectively suppresses the cholesteric twist at high coverage.

To probe how the relative contributions of the two acyl chain geometries influence anchoring, we studied mixed DLPC:DOPC systems across a range of total lipid concentrations (Fig. 2). All images were acquired near the center of the TEM grid to minimize variations in film geometry. 

\begin{figure}[H]
    \centering
    
    \captionsetup{font=scriptsize} 
    \includegraphics[width=0.9\linewidth]{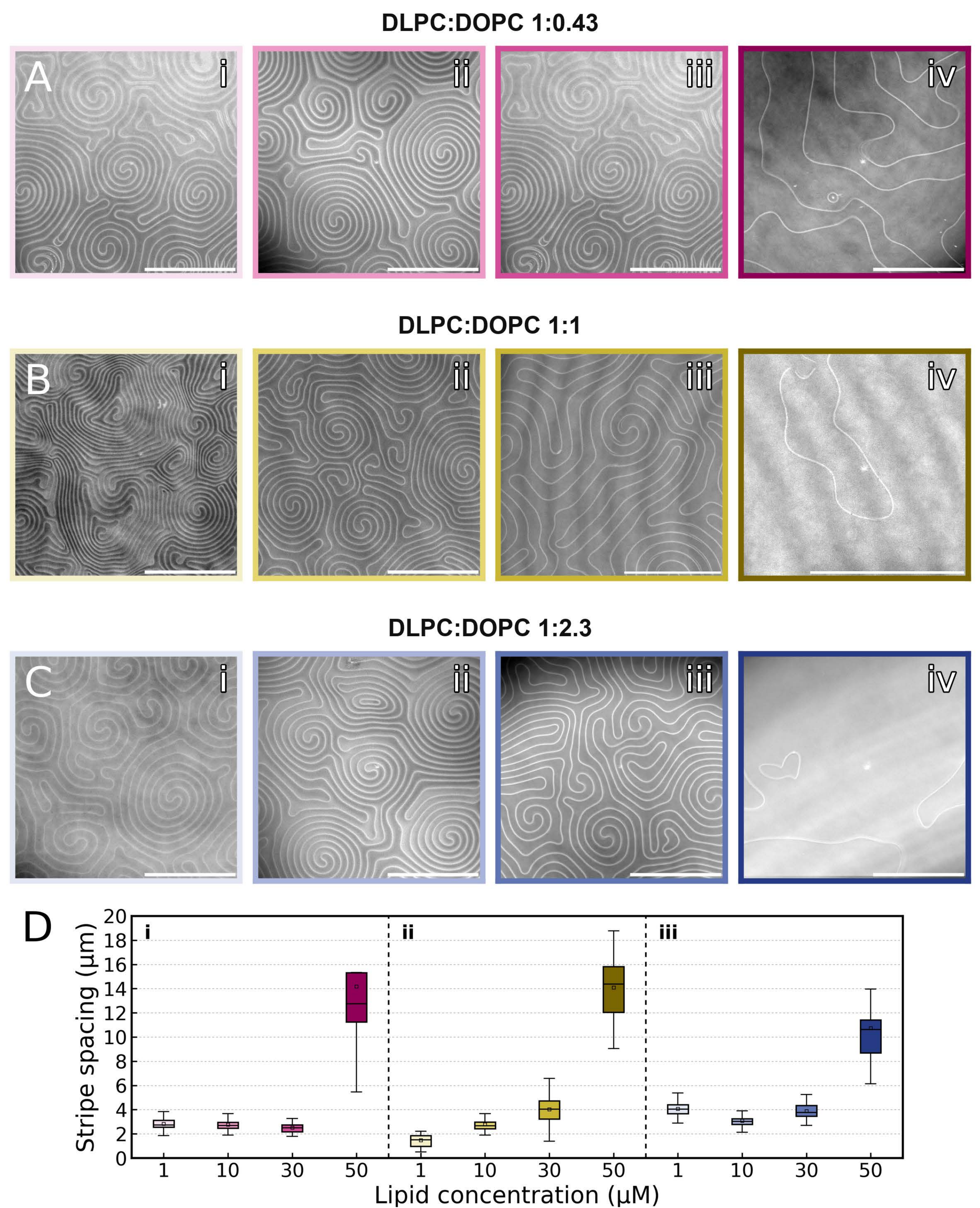}
    \caption{Cholesteric interfacial stripe patterns as a function of DLPC:DOPC mixing
    ratio and concentration. (A) DLPC: DOPC 1:0.43, (B) DLPC: DOPC 1:1, (C) DLPC: DOPC 1:2.3. Confocal z-stacks of representative textures of cholesterics at the liquid crystal-water interface under increasing lipid concentrations (i-1, ii-10, iii-30, iv-50 µM). (D) Plots of the stripe spacing against overall lipid concentration for i) DLPC: DOPC 1:0.43, ii) DLPC: DOPC 1:1, and iii) DLPC: DOPC 1:2.3. The box colors surrounding the images correspond to the colors used in the plots for visual consistency. (Scale bar: 50 µm.)}
    \label{fig:2}
\end{figure}

The DLPC-rich 1:0.43 mixture displays highly ordered stripes at 1~µM that remain well defined at 10 and 30~µM (Fig.~2A). At 50~µM, the stripe contrast decreases and large, irregular homeotropic regions emerge. Correspondingly, the characteristic stripe spacing decreases from 1 to 30~µM before increasing at 50~µM (Fig.~2D-i).

For the equimolar DLPC:DOPC mixture, the stripes observed at 1~µM are less regular than those of the DLPC-rich system (Fig. 2B). Note that the larger-scale intensity modulations at 10-50~µM arise from maximum-intensity projections through interfaces with uneven heights and do not represent cholesteric periodicity. The measured spacing is smaller than the expected cholesteric half-pitch at 1~µM, becomes more uniform and approaches $p/2\sim2.5$~µm at 10~µM, and broadens as the homeotropic regions expand at 30~µM (Fig.~2D-ii). At 50~µM, the periodic texture fades and large homeotropic regions dominate the interface. In the DOPC-rich 1:2.3 mixture, the larger spacing at 1~µM is consistent with more extensive homeotropic regions (Fig.~2C). The spacing changes only weakly between 10 and 30~µM, whereas at 50~µM the stripe pattern largely disappears, leaving a predominantly homeotropic interface with isolated twist-disclination lines \cite{tran2017change,tran2020swelling}.

These results show that the mixed systems do not simply interpolate between pure DLPC and DOPC. The DLPC-rich mixture maintains the most regular stripes over the widest concentration range, whereas increasing the DOPC fraction generally produces greater spatial heterogeneity and more extensive homeotropic regions. Lipid composition therefore influences both the magnitude and spatial uniformity of the anchoring imposed at the interface.

We interpret these trends as reflecting a balance between tail-mediated local anchoring and packing-mediated collective anchoring. The longer C18 chains of DOPC may promote strong local homeotropic alignment at low surface coverage, accounting for the early appearance of enlarged homeotropic regions in DOPC-containing mixtures. However, their kinked \textit{cis}-unsaturated conformations are expected to hinder dense, uniform packing and produce spatial variations in effective anchoring. In contrast, the shorter, saturated DLPC chains may impose weaker local anchoring at low coverage but support more uniform collective anchoring as surface density increases.

The composition-dependent textures provide indirect support for this molecular interpretation. Efficient DLPC packing is associated with more uniform collective anchoring, whereas the longer, kinked DOPC chains are associated with locally strong but spatially heterogeneous anchoring. Although lipid packing, composition, and chain conformation were not directly resolved, the cholesteric texture reports their influence through measurable changes in director configuration and stripe morphology.

We also considered whether intrinsic lipid-lipid phase separation could contribute to the observed heterogeneity. Both DLPC (12:0) and DOPC (18:1) are in the fluid, liquid-disordered state at room temperature \cite{bag2014temperature,kim2026curvature}, and binary mixtures of fluid phosphatidylcholines without cholesterol are generally expected to remain miscible. Liquid-ordered/liquid-disordered phase separation more commonly arises when cholesterol promotes acyl-chain ordering in mixtures containing a high-melting saturated lipid \cite{veatch2003separation,harayama2018understanding}. At the present liquid crystal-aqueous interface, the cholesteric may provide a different source of cooperative coupling, as lipid organization modifies surface anchoring while the resulting director distortion carries an elastic cost. This coupling could potentially promote lateral redistribution or packing heterogeneity within an otherwise miscible lipid layer, an intriguing possibility for future study. However, because the present probes are not species-specific, we cannot determine whether the observed regions reflect local changes in DLPC/DOPC composition or differences in molecular packing. We therefore describe these features as lipid-enriched regions or interfacial heterogeneity rather than distinct compositional phases.

\subsection{Pitch-dependence of  lipid-induced anchoring of cholesterics}

\indent \indent To examine how intrinsic twist elasticity influences lipid-induced anchoring, we varied the cholesteric pitch by preparing mixtures containing 2.8, 5, and 10 wt.-\% CB15 (Fig.~3, rows i-iii), corresponding to pitches of approximately 5, 1, and 0.6~µm, respectively \cite{tran2018shaping}. Each cholesteric was exposed to DLPC (Fig.~3A), DOPC (Fig.~3E), or mixed DLPC:DOPC solutions (Fig.~3B-D) at a fixed total lipid concentration of 10~µM.

In a cholesteric liquid crystal, the director field $\mathbf{n}$ rotates with a preferred helical pitch $p$, corresponding to a spontaneous twist wavenumber $q_0=2\pi/p$ \cite{de1993physics}. The twist contribution to the Frank free energy density can be written as $f_{\mathrm{twist}}\sim(K_2/2)(q-q_0)^2$, where $K_2$ is the twist elastic constant and $q$ is the local twist wavenumber imposed by the actual director configuration. In the undistorted cholesteric state, $q=q_0$, so the preferred helical structure does not itself carry an elastic penalty. However, when lipid adsorption imposes a homeotropic easy axis at the interface, the director near the surface is forced away from its preferred cholesteric configuration, such that the local twist can deviate from $q_0$ \cite{kiselev2005twist}. This distortion produces an elastic cost that scales with $(\delta q)^2$, where $\delta q=q-q_0$.

This scaling shows why pitch provides a direct handle on the balance between bulk elasticity and surface anchoring. If a deformation changes the apparent pitch by an amount $\Delta p$, then the corresponding change in wavenumber is approximately $\delta q \approx -(2\pi/p^2)\Delta p$. Thus, for a comparable pitch distortion, shorter-pitch cholesterics experience a larger change in twist wavenumber and therefore a larger elastic penalty. Equivalently, complete local unwinding of the helix corresponds to changing the twist from $q_0$ toward $q\approx0$, giving an elastic cost that scales as $K_2q_0^2/2\propto K_2/p^2$. Shortening the pitch therefore increases the energetic cost of suppressing or strongly perturbing the helix.

The surface contribution can be described qualitatively by an anchoring energy density $f_{\mathrm{s}}\sim(W/2)\sin^2\theta$, where $W$ is the effective lipid-induced anchoring coefficient and $\theta$ is the angle between the local director and the lipid-imposed easy axis. The observed interfacial texture is therefore determined by the competition between the bulk twist penalty, set by $K_2$ and $p$, and the surface anchoring tendency, set by $W$. When the anchoring energy is sufficiently strong relative to the twist penalty, the interface can drive larger homeotropic regions or partial helix unwinding. When the pitch is shorter, the larger twist penalty resists this distortion, favoring periodic fingerprint textures that preserve more of the intrinsic cholesteric twist. This framework is consistent with prior studies showing that cholesteric stripe patterns, helix unwinding, pitch-jump transitions, and confinement-dependent defect textures are governed by the balance between surface anchoring, intrinsic pitch, and confinement geometry \cite{de1993physics, kiselev2005twist, mckay2013unwinding, guo2016cholesteric, krakhalev2019orientational, lavrentovich2020undulation, tran2017change, tran2018shaping}.

At 2.8 wt.-\% CB15, all lipid compositions produce visible fingerprint textures with the largest stripe spacings observed in the series (Fig.~3A-E, row i; Fig.~3F). Note again that the larger-scale intensity variations in the micrographs arise from uneven interfacial height and confocal $z$-projection and do not represent cholesteric periodicity. DLPC-containing systems exhibit relatively continuous stripes with regular spacing and orientation, whereas increasing DOPC content produces more distorted textures. Pure DOPC shows the largest median spacing and local disruptions in stripe continuity (Fig.~3E, row i). Although the DOPC distributions have a larger absolute spread, their relative variation is comparable to that of the DLPC systems (coefficient of variation, CV $\approx20\%$), reflecting their larger mean spacing rather than a distinctly broader normalized distribution.

These long-pitch textures indicate that the cholesteric helix can accommodate lipid-induced anchoring while retaining periodic order. Because the elastic penalty for suppressing the preferred twist scales as $K_2/p^2$, the $p\approx5$~µm cholesteric is expected to resist anchoring-induced distortion less strongly than the shorter-pitch systems, assuming only modest changes in $K_2$. The remaining morphological differences therefore likely reflect variations in the local anchoring imposed by the lipid layer. DLPC-containing interfaces preserve the cholesteric periodicity more uniformly, whereas DOPC may produce stronger local homeotropic anchoring through its longer tails together with more spatially heterogeneous organization arising from its kinked unsaturated chains.

At 5 wt.-\% CB15, the decrease in pitch produces denser fingerprint textures and smaller median stripe spacings for every lipid composition (Fig.~3A-E, row ii; Fig.~3F). DLPC retains well-defined stripes, while the mixed systems become more spatially variable as the DOPC fraction increases. In the DOPC-rich mixture, lipid-enriched regions coexist with areas containing periodic stripes (Fig.~3D, row ii), and pure DOPC remains striped but more distorted than the DLPC-rich systems (Fig.~3E, row ii). The normalized distribution widths overlap across compositions (CV $\approx19$-32\%). Thus, no systematic difference in distribution width between pure DLPC and pure DOPC is resolved at this pitch.

\begin{figure}[H]
\centering
\captionsetup{font=scriptsize}
\includegraphics[width=1\linewidth]{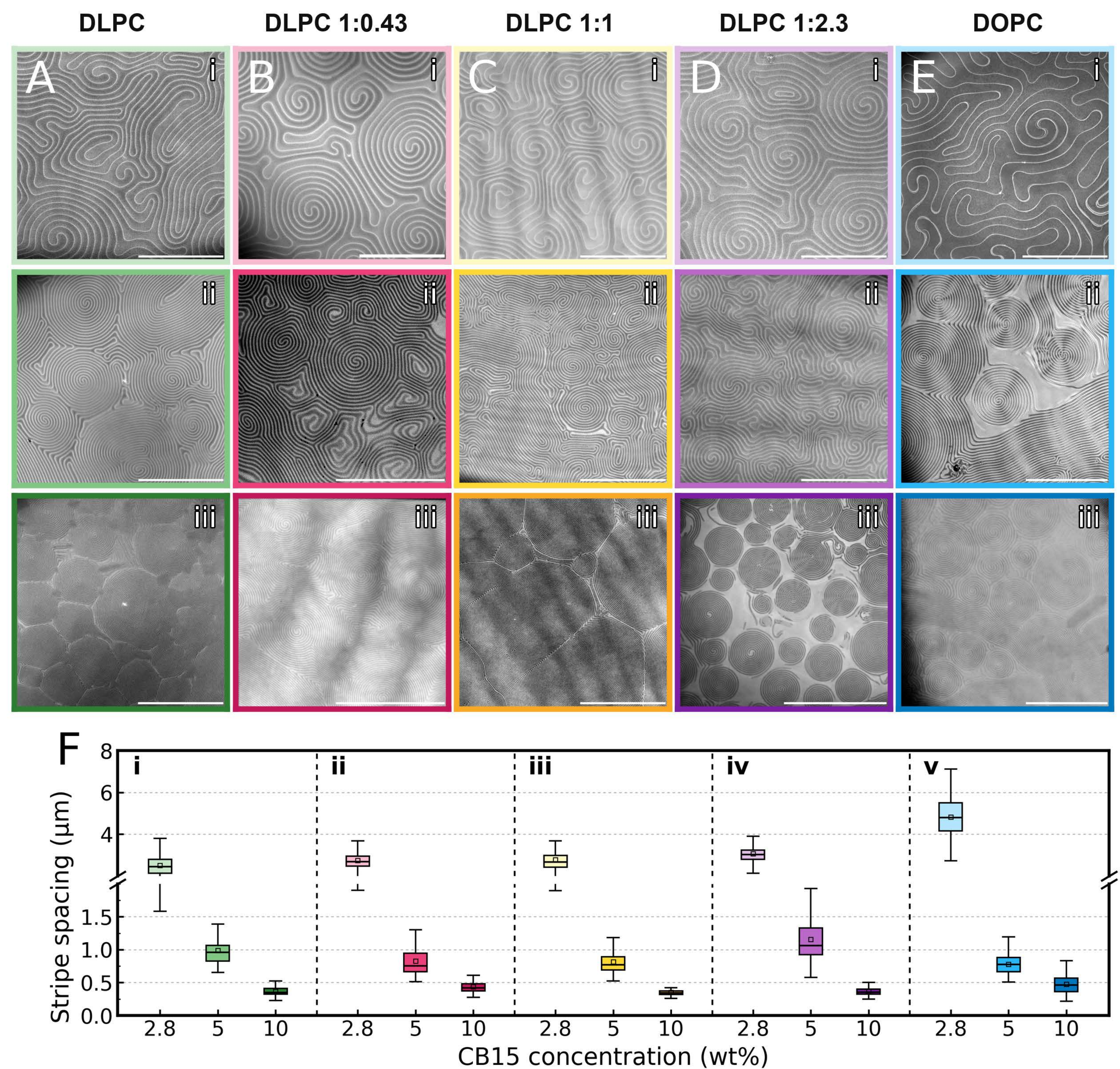}
\caption{Cholesteric interfacial stripe morphology and stripe spacing as functions of chiral dopant concentration and lipid composition. (A-E) Representative confocal microscopy images of cholesteric textures at the liquid crystal-water interface for different lipid compositions: (A) DLPC, (B) DLPC 1:0.43, (C) DLPC 1:1, (D) DLPC 1:2.3, and (E) DOPC. Rows i through iii correspond to 2.8, 5, and 10 wt.-\% CB15, respectively, and therefore to decreasing cholesteric pitch. (F) Stripe-spacing distributions for each lipid composition and CB15 concentration. Box colors correspond to the borders around the representative micrographs: DLPC (green), DLPC 1:0.43 (pink), DLPC 1:1 (yellow), DLPC 1:2.3 (purple), and DOPC (blue). Measurements include only regions containing identifiable periodic stripes ($n\approx75$--126 peak-to-peak measurements per box, from 5-6 samples; tests are across measurements). Significance was assessed with the Kruskal-Wallis test across CB15 concentrations ($p<0.05$ throughout) and the Mann-Whitney $U$ test between compositions (see text). The corresponding single-lipid image series and grayscale intensity profiles are provided in Figs.~S5 and S6. Scale bar: 50 µm.}
\label{fig:3}
\end{figure}

At 10 wt.-\% CB15, the regions retaining fingerprint textures exhibit the smallest stripe spacings (Fig.~3A-E, row iii). DLPC and the DLPC-rich mixtures retain comparatively continuous stripes, whereas the DOPC-rich mixture contains large lipid-enriched regions that interrupt the periodic texture (Fig.~3D, row iii). Pure DOPC also retains stripes but with greater spatial variation in continuity and spacing (Fig.~3E, row iii). Differences in median spacing among lipid compositions become less pronounced at this shortest pitch, although pure DOPC exhibits a broader relative distribution and more heterogeneous morphologies.

The systematic decrease in stripe spacing with CB15 concentration confirms that the intrinsic pitch primarily controls periodicity wherever the fingerprint texture remains intact. Increasing CB15 concentration reduces the stripe spacing for every lipid composition (Kruskal-Wallis test, $p<0.05$ for each composition)\cite{virtanen2020scipy,li2026cholestericstats}, and the median spacing differs between pure DLPC and pure DOPC at 2.8 and 5 wt.-\% CB15 (Mann-Whitney $U$ test, $p<0.05$), and only slightly at 10 wt.-\%.

Interestingly, differences in relative distribution width are limited at the longer pitches, with comparable coefficients of variation across compositions at 2.8 and 5 wt.-\% CB15. Pure DOPC becomes distinctly broader only at 10 wt.-\% CB15. This trend may reflect the interaction between increased twist elasticity and a spatially heterogeneous anchoring field. At shorter pitch, the larger energetic cost of smoothly distorting or locally unwinding the helix constrains regions with relatively uniform anchoring to remain close to the intrinsic periodicity. In DOPC-rich systems, however, spatial variations in anchoring may instead be accommodated through localized stripe distortions and coexistence with lipid-enriched regions, producing more distorted stripe morphologies. Thus, the increased twist penalty with decreasing pitch may enhance the morphological consequences of heterogeneous interfacial anchoring.

Together, these results indicate that pitch sets the characteristic stripe spacing, whereas lipid composition controls how regularly that pitch is expressed across the interface. DLPC-rich systems generally preserve a more regular alternation of planar and homeotropic regions, consistent with less spatial distortion of the cholesteric pitch. DOPC-containing systems can produce stronger local homeotropic anchoring but also more heterogeneous interfacial organization, resulting in distorted stripes and lipid-enriched regions that coexist with periodically ordered areas.

\subsection{Coexisting textures under variable confinement}

\indent \indent The pronounced heterogeneity observed for the DOPC-rich mixture at the shortest pitch motivated a closer examination of this condition. The sample contained DLPC:DOPC = 1:2.3 at a total lipid concentration of 10~µM and 10 wt.-\% CB15 (Fig.~3D, row iii). Because the TEM-grid geometry produces spatial variations in film thickness within a continuous specimen, it provides a qualitative means of examining the influence of confinement while lipid composition, concentration, and cholesteric pitch remain fixed. 

\begin{figure}[H]
\centering
\captionsetup{font=scriptsize}
\includegraphics[width=0.75\linewidth]{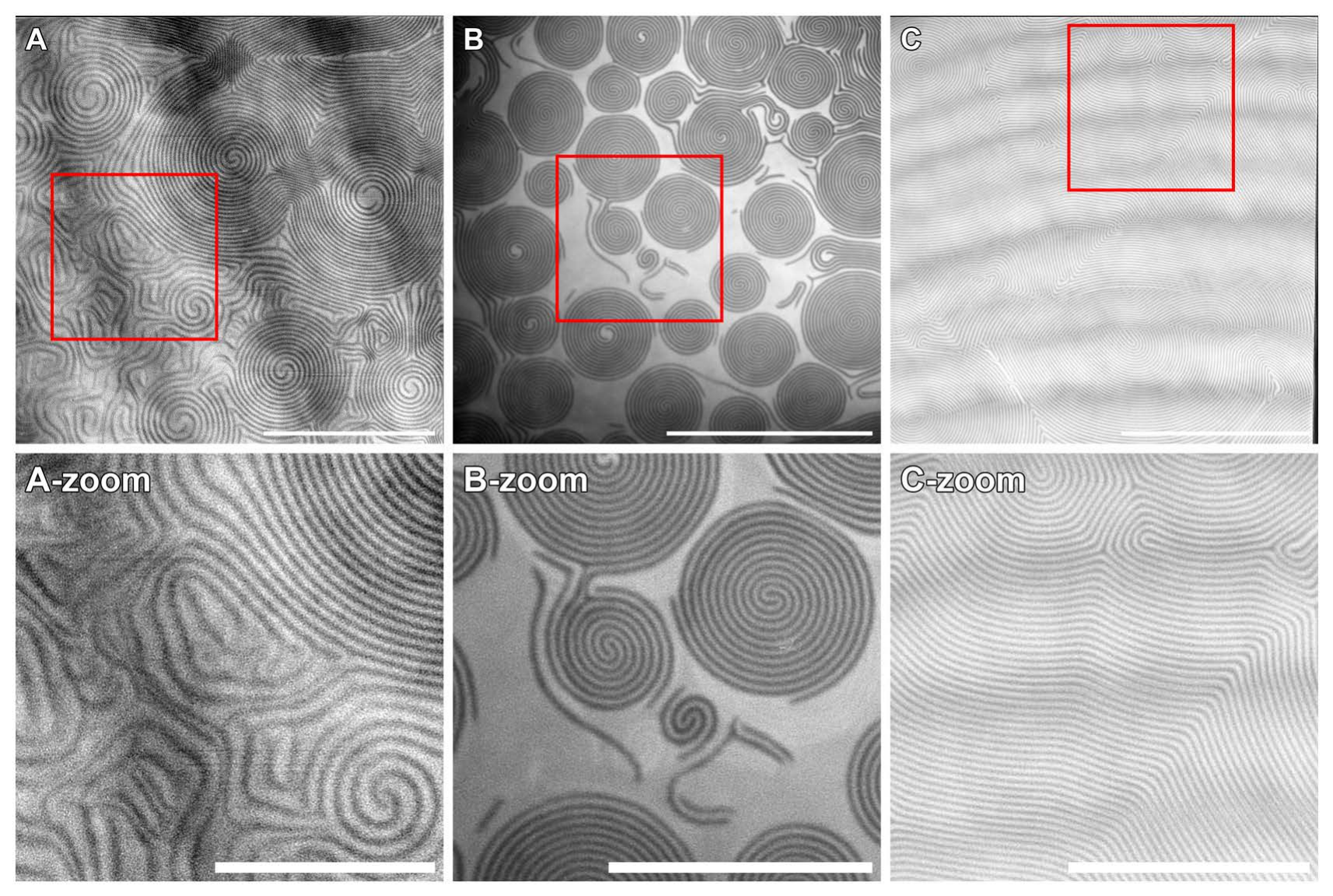}
\caption{Coexisting interfacial textures in a DOPC-rich mixed-lipid cholesteric containing DLPC:DOPC = 1:2.3, 10~µM total lipid, and 10 wt.-\% CB15. Representative confocal micrographs were acquired from different regions of the same TEM-grid specimen. (A) Distorted fingerprint texture near a grid edge. (B) Broad lipid-enriched boundaries surrounding domains with closely spaced fingerprint stripes. (C) Large regions of comparatively regular fingerprint stripes. Scale bars: 50~µm; zoomed image: 25~µm.}
\label{fig:4}
\end{figure}

Distinct interfacial textures coexist within the same sample (Fig.~4). Fig.~4A shows continuous but locally distorted fingerprint stripes, including undulated and nearly perpendicular orientations that suggest spatial variations in the pitch-axis direction \cite{tran2017change,lavrentovich2020undulation,blanc2023helfrich}. In Fig.~4B, broad fluorescent boundaries surround domains whose interiors retain closely spaced fingerprint stripes. Fig.~4C instead shows extended regions of comparatively regular stripes with more uniform fluorescence and without the broad bright boundary features associated with the lipid-enriched regions in Fig.~4B. Note that the broad intensity modulations in Figs.~4A and 4C arising from uneven interfacial height and $z$-projection are distinguishable from the finer cholesteric stripe periodicity.

Because the chemical conditions are uniform across the specimen, this texture coexistence is most plausibly associated with local variations in confinement. At 10 wt.-\% CB15, the short-pitch cholesteric strongly resists local unwinding because the twist penalty scales approximately as $K_2/p^2$. Film thickness determines how much of the preferred helical rotation can be accommodated between the confining boundaries and therefore modifies the balance between intrinsic twist elasticity and lipid-induced homeotropic anchoring. Different local thicknesses may consequently favor regular fingerprint stripes, distorted textures, or regions in which the periodic pattern is interrupted. The local thickness was not measured, however, so a specific texture cannot be assigned to a particular confinement regime.

The broad, bright domains in Fig.~4B are described as lipid-enriched regions based on their fluorescence morphology and the complementary imaging controls discussed in the Supporting Information. Their coexistence with periodically ordered finger-print domains suggests that lateral variations in interfacial organization redistribute the effective anchoring conditions without eliminating cholesteric order throughout the sample. The interface may accommodate frustration by preserving the short-pitch helix within stripe-forming regions while concentrating stronger anchoring or altered lipid organization at neighboring domains.

These observations identify confinement as an additional factor governing pattern selection. Under fixed chemical conditions, variations in film geometry coincide with the coexistence of regular and distorted fingerprint textures and lipid-enriched boundary regions. Quantitatively relating these states to confinement will require samples with controlled thickness, together with direct measurements of local film geometry and lipid organization.

\subsection{Interfacial lipid mobility measured by FRAP}

\begin{figure}[H]
    \centering
    
    \captionsetup{font=scriptsize}
    \centering
    \includegraphics[width=1\linewidth]{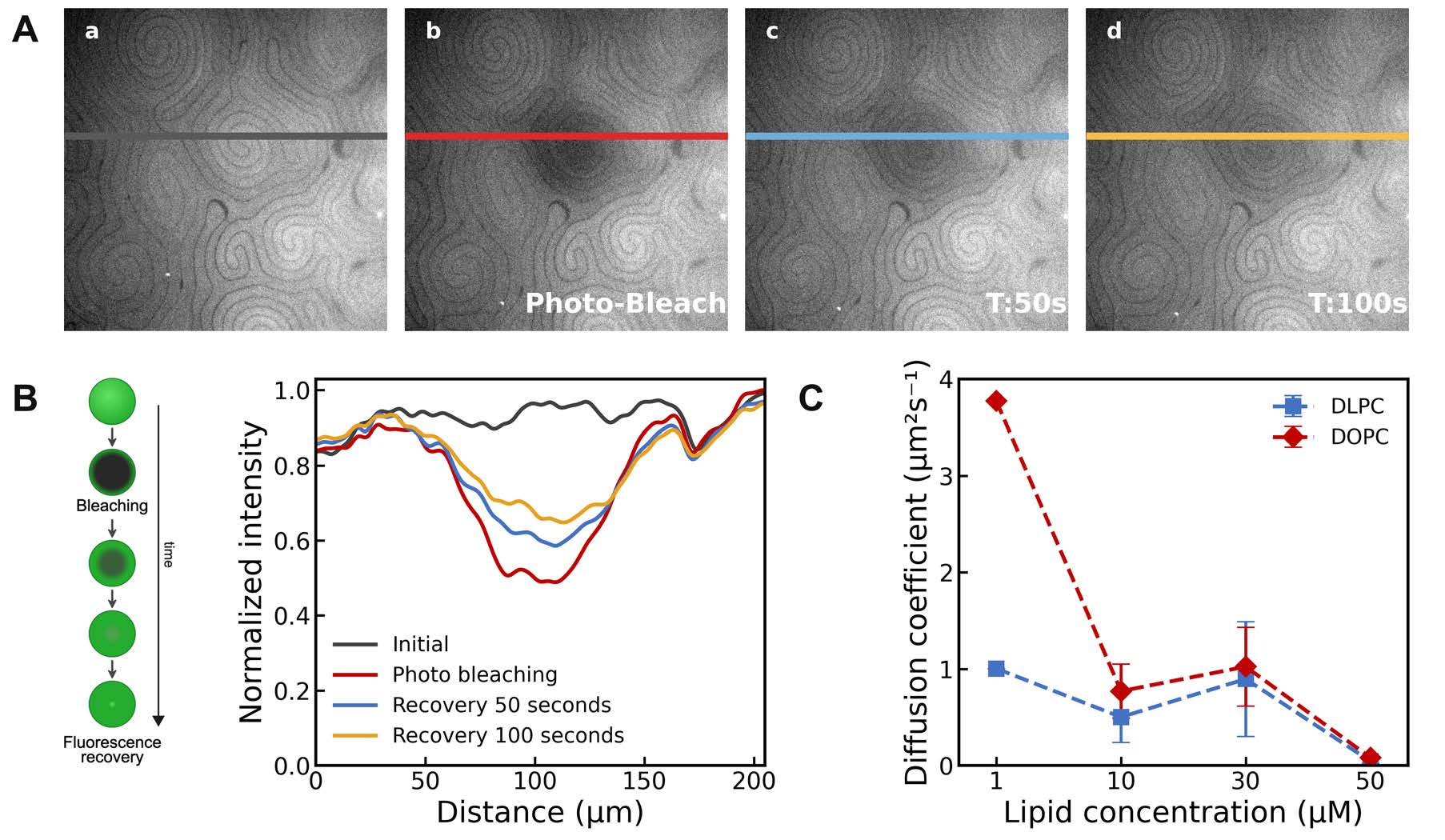}
  \caption{Fluorescence recovery after photobleaching (FRAP) at the cholesteric-aqueous interface. (A) Representative measurement for an interface containing 10~µM DLPC and displaying a fingerprint texture. Images show the fluorescence before bleaching (a), immediately after bleaching (b), and after 50~s (c) and 100~s (d) of recovery. (B) Schematic of the bleach-and-recovery sequence (left) and the corresponding fluorescence intensity profiles along the bleaching line (right). (C) Interfacial diffusion coefficients of Texas Red-DHPE in DLPC- and DOPC-containing samples, determined using the J\"{o}nsson analysis method \cite{jonsson2008method}. A representative normalized recovery curve is shown in Fig.~\ref{fig:s8}. Scale bar: 50~µm.}
    \label{fig:5}
\end{figure}

\indent \indent Fluorescence recovery after photobleaching (FRAP) was used to probe the lateral mobility of Texas Red-DHPE at the cholesteric-aqueous interface (Fig.~5) \cite{brake2003active,loren2015fluorescence,day2012analysis,van2008membrane,vanni2014sub,cumberland2018bending,akartuna2008stabilization}. Measurements were performed by wide-field fluorescence microscopy through the aqueous phase after excess vesicles were removed from solution (see Materials and Methods). This imaging geometry also avoids the fluorescence contrast inversion associated with confocal imaging through the cholesteric. Because the measured diffusion coefficients correspond directly to Texas Red-DHPE, we refer to them as fluorescent-probe mobilities, as they are not direct measurements of DLPC or DOPC molecular diffusion.

Fig.~5A illustrates a representative FRAP measurement for an interface containing 10~µM DLPC. Fluorescence decreases sharply upon bleaching and subsequently recovers as fluorescent probes redistribute into the bleached region (Fig.~5B). The corresponding normalized recovery curve, together with the simFRAP analysis used to illustrate the recovery behavior, is provided in Fig.~\ref{fig:s8}.

The diffusion coefficients reported in Fig.~5C were determined independently using a MATLAB implementation of the method described by J\"{o}nsson et al. \cite{jonsson2008method}. Rather than fitting the mean fluorescence recovery curve, this method analyzes the spatial evolution of the bleached intensity profile using a Hankel-transform-based approach to determine lateral diffusion. Thus, the diffusion coefficients in Fig.~5C are not obtained from the representative recovery curve shown in Fig.~\ref{fig:s8}.

The measured fluorescent-probe diffusion coefficients vary with lipid concentration (Fig.~5C). At 1~µM, DOPC exhibits a higher apparent diffusion coefficient ($D_{\mathrm{DOPC}}\approx3.8~\text{\textmu}\mathrm{m}^2\,\mathrm{s}^{-1}$) than DLPC ($D_{\mathrm{DLPC}}\approx1.0~\text{\textmu}\mathrm{m}^2\,\mathrm{s}^{-1}$). The measured values are consistent with the greater conformational freedom and less cohesive organization expected for dilute DOPC layers. At 10~µM, the diffusion coefficients decrease to approximately $0.5~\text{\textmu}\mathrm{m}^2\,\mathrm{s}^{-1}$ for DLPC and $0.8~\text{\textmu}\mathrm{m}^2\,\mathrm{s}^{-1}$ for DOPC, while at 30~µM both are approximately $1.0~\text{\textmu}\mathrm{m}^2\,\mathrm{s}^{-1}$. The replicate distributions overlap at both concentrations, and therefore no clear difference between DLPC and DOPC is resolved in this intermediate concentration range.

At 50~µM, measurable recovery is strongly suppressed for both lipid systems, indicating highly restricted lateral probe mobility. Taken together, the measurements show an overall shift from relatively mobile interfaces at low lipid concentration toward increasingly constrained dynamics at high concentration. The absence of a resolved DLPC-DOPC difference at intermediate concentrations, together with the strongly suppressed recovery for both systems at 50~µM, suggests that collective interfacial organization becomes increasingly important relative to differences associated with the individual acyl chain structures as surface coverage increases.

These mobility measurements complement the optical textures by showing that the expansion of homeotropically anchored regions is accompanied by increasingly restricted lateral motion of the fluorescent phospholipid probe. The combined FRAP and imaging results therefore identify surface coverage as a common control parameter linking interfacial mobility to the macroscopic anchoring response of the cholesteric. Although the measurements do not establish that reduced mobility directly causes stronger anchoring, they demonstrate that the transition toward predominantly homeotropic alignment occurs together with the formation of a collectively organized, dynamically constrained lipid interface.

\section{Conclusion}

\indent \indent This study demonstrates that lipid molecular structure, surface coverage, cholesteric pitch, and confinement jointly regulate anchoring at cholesteric liquid crystal-aqueous interfaces. Comparison of saturated DLPC and unsaturated DOPC shows that the transition from planar alignment, through fingerprint textures, toward predominantly homeotropic alignment cannot be explained by hydrocarbon chain length alone. Instead, the observed behavior is consistent with a balance between local lipid-liquid crystal interactions and collective properties of the interfacial layer, including packing, mobility, and spatial uniformity \cite{israelachvili2011intermolecular,nagle2000structure}.

DLPC-containing interfaces generally produce more regular fingerprint morphologies and more readily approach predominantly homeotropic alignment at high surface coverage. This behavior is consistent with the shorter, saturated chains of DLPC supporting relatively uniform collective anchoring as the interfacial density increases. DOPC-containing interfaces exhibit locally enlarged homeotropic regions, distorted stripes, and interruptions of the periodic texture. The longer DOPC chains may promote strong local interactions with the liquid crystal, while their \textit{cis}-double-bond kinks hinder uniform lateral organization \cite{pinot2014polyunsaturated,brake2005formation}. The mixed-lipid systems do not behave as simple interpolations between the pure lipids, further showing that anchoring depends on the collective organization of both molecular species.

Variation of the CB15 concentration shows that the intrinsic pitch primarily determines the stripe spacing wherever a periodic fingerprint texture remains intact. Increasing CB15 concentration shortens the pitch and decreases the measured stripe spacing across all lipid compositions, consistent with the increasing elastic cost of locally perturbing or unwinding the helix. Lipid composition instead controls how regularly this intrinsic periodicity is expressed. Relative distribution widths are comparable across compositions at 2.8 and 5 wt.-\% CB15, whereas pure DOPC becomes distinctly broader at 10 wt.-\% CB15. This behavior is consistent with the larger short-pitch twist penalty enhancing the morphological consequences of a spatially heterogeneous anchoring field.

Variable confinement introduces an additional source of texture selection. Within a chemically uniform DOPC-rich, short-pitch specimen, regular fingerprint stripes coexist with distorted textures and lipid-enriched boundary regions. These variations are most plausibly associated with local differences in film geometry, which alter the amount of helical rotation that can be accommodated between the confining boundaries. Because the local thickness was not measurable, however, individual textures cannot yet be assigned to specific confinement regimes.

FRAP provides complementary information on interfacial fluorescent-probe mobility \cite{loren2015fluorescence,day2012analysis}. The measurements show an overall shift toward more restricted lateral dynamics as lipid concentration increases. The apparently higher DOPC probe mobility at 1~µM is suggestive of an influence of acyl chain structure under dilute conditions. At intermediate concentrations, the DLPC- and DOPC-containing samples show overlapping diffusion coefficients within their experimental scatter, while measurable recovery becomes strongly suppressed for both systems at the highest concentration. Together with the optical textures, these results show that increasing surface coverage is accompanied by both stronger homeotropic anchoring and increasingly constrained interfacial dynamics.

Because local packing density, lipid composition, and molecular conformation were not measured directly, the proposed mechanism remains an interpretation based on cholesteric texture analysis, complementary fluorescence imaging, and FRAP. Nevertheless, the cholesteric provides a sensitive reporter of interfacial organization by amplifying molecular-scale changes into measurable differences in director configuration, stripe morphology, spacing, and interfacial mobility. The fluorescence contrast observed through the cholesteric further demonstrates that its anisotropic director field can transduce local orientational structure into an optical signal, providing an additional readout of the interfacial state. Taken together, the results suggest that acyl chain structure regulates anchoring through both direct interactions with the liquid crystal and its influence on the collective organization and spatial heterogeneity of the lipid layer.

Future studies using controlled film thicknesses, such as microfabricated wells, could establish quantitative relationships among confinement, pitch, and texture selection \cite{guo2016cholesteric}. Direct characterization by polarization-resolved fluorescence, Brewster angle microscopy, or interface-specific vibrational spectroscopy could test whether the lipid-enriched regions reflect variations in packing density, molecular conformation, or local composition \cite{chen2010interfacial,fellows2024spiral}. Microfluidic interfaces could further control lipid delivery and adsorption kinetics while allowing anchoring transitions to be monitored in real time \cite{dedeoglu2026soft}.

Collectively, these findings show that cholesteric interfacial organization emerges from the coupling of lipid structure, elastic frustration, and geometric confinement. This framework provides a basis for controlling anchoring strength and texture morphology in responsive liquid crystal interfaces for sensing and bioinspired soft materials \cite{gupta1997optical,bisoyi2021liquid,ren2025stretchable}.

\newpage
\section{Materials and Methods}
\subsection{Materials}
\begin{sloppypar}
\indent \indent The lipids 1,2-dilauroyl-sn-glycero-3-phosphocholine (DLPC) and 1,2-dioleoyl-sn-glycero-3-phosphocholine (DOPC) were obtained from Avanti Polar Lipids. For fluorescence imaging, the lipid mixtures were labeled with 1 mole percent (mol \%) of the fluorescent probe Texas Red™ 1,2-dihexadecanoyl-sn-glycero-3-phosphoethanolamine, triethylammonium salt (Texas Red™ DHPE, Thermo Fisher Scientific). For the two-color fluorescence control (Fig.~S7), Texas Red-DHPE and \textit{N}-(fluorescein-\hspace{0pt}5-\hspace{0pt}thiocarbamoyl)-\hspace{0pt}1,2-\hspace{0pt}dihexadecanoyl-\hspace{0pt}sn-\hspace{0pt}glycero-\hspace{0pt}3-\hspace{0pt}phosphoethanolamine, triethylammonium salt (Fluorescein DHPE, hereafter FITC-DHPE, Thermo Fisher Scientific) were used as spectrally distinct DHPE-based fluorescent probes. Because neither probe is chemically specific to DLPC or DOPC, the two-color experiment was used to test whether the interfacial stripe pattern was reproduced in an independent fluorescence channel rather than to distinguish the two phosphatidylcholine species. The aqueous hydration solution (10 mM Tris, 0.1 M NaCl, pH 9.68) was prepared using Tris buffer powder obtained from Sigma-Aldrich. For the cholesteric liquid crystal phase, 4-cyano-4$'$-pentylbiphenyl (5CB) was purchased from TCI Chemicals and doped with the chiral agent (S)-4-cyano-4$'$-(2-methylbutyl)biphenyl (CB15, Thermo Fisher Scientific) to induce a cholesteric pitch. PELCO® center-marked copper grids (50 mesh, 3.0 mm outer diameter, Ted Pella, Inc.) were used as sample supports. To render the substrates hydrophobic and control the surface anchoring of the liquid crystal, glass surfaces were treated with trichloro(octadecyl)silane (OTS, Sigma-Aldrich).
\end{sloppypar} 

\subsection{Liposome Preparation }
\indent \indent Lipid vesicles were prepared using the standard thin-film hydration method. The primary lipids (DLPC or DOPC) and 1 mol-\% of the fluorescent probe Texas Red™ DHPE were co-dissolved in chloroform in a glass vial. The organic solvent was evaporated under a gentle stream of nitrogen gas to form a uniform thin lipid film on the vial wall. To ensure the complete removal of residual organic solvent, the lipid film was kept under vacuum overnight. The dried film was then hydrated with a 10 mM Tris buffer containing 0.1 M NaCl (pH 9.68) at room temperature, which is well above the gel-to-liquid crystalline phase transition temperature of the lipids. The mixture was vigorously vortexed to yield a suspension of multilamellar vesicles (MLVs). To obtain uniform unilamellar vesicles, the suspension was first subjected to bath sonication for 10 min and subsequently extruded through 100-nm and 50-nm polycarbonate membranes using a mini-extruder (Avanti Polar Lipids).

After extrusion, the hydrodynamic size distribution of the lipid vesicles was characterized by dynamic light scattering (DLS). For the DLPC vesicles, DLS measurements were performed after storage at 4 °C for up to 5 days to assess the stability of the vesicle dispersion (Figure S8).

\subsection{Experimental Setup}
\indent \indent TEM-grid-supported liquid crystal films were prepared using a configuration adapted from previously reported liquid crystal-aqueous interface systems\cite{Brake2003BiomolecularInteractions}. Glass substrates were functionalized with octadecyltrichlorosilane (OTS) to promote homeotropic anchoring of the liquid crystal at the solid substrate. A Grace Bio-Labs SecureSeal™ imaging spacer with an inner diameter of 9 mm and a thickness of 0.12 mm was adhered to the OTS-functionalized glass substrate to define the aqueous sample well. A TEM grid was placed at the center of the well, and a small volume of liquid crystal was deposited onto the grid. Excess liquid crystal was removed using a glass capillary tube with a capacity of 5-25 µL, leaving liquid crystal confined within the apertures of the TEM grid.

Lipid vesicle suspensions were then introduced into the well to establish contact with the liquid crystal-aqueous interface. The samples were incubated for 30 min to allow vesicle adsorption, rupture, and lipid assembly at the interface. After incubation, the surrounding solution was gently exchanged with fresh Tris buffer (10 mM Tris, 0.1 M NaCl, pH 9.68) to remove excess vesicles remaining in the aqueous phase while minimizing disturbance to the interfacial lipid layer. The well was subsequently covered with a clean coverslip to reduce evaporation during imaging. The assembled samples were then transferred to a microscope for observation.

\subsection{Optical Characterization}
\indent \indent The main imaging system used in the experiments was a Leica TCS SP8 stimulated emission depletion (STED) super-resolution confocal microscope, mounted on a DMI6000 B inverted stand and equipped with an HC PL APO CS2 93×/1.30 glycerol-immersion objective. A tunable white light laser was used for the excitation of Texas Red™ DHPE, and a 660 nm STED laser was applied for emission depletion to achieve super-resolution. To avoid scattering from the depletion laser, the fluorescent emission was carefully collected within a restricted wavelength window from 604.5 to 635.4 nm. Additionally, time-gated detection (0.3-6.5 ns) was applied to further suppress background signal and enhance STED resolution. The pinhole was set to 1 Airy unit. Z-stacks were obtained by a motorized focus control with a z-step size of 0.26 µm. Acquired images were deconvolved using SVI Huygens software, and ImageJ was used for the 2D projections of the confocal z-stacks. Raw 16-bit micrographs were converted to publication-format images with calibrated scale bars using \texttt{micrograph-batch}, an open-source command-line tool written for this work \cite{li2026micrographbatch}; all crop, brightness, contrast, and scale-bar parameters are recorded in a CSV configuration file, so that every displayed micrograph can be regenerated from the raw data. For the two-color co-localization control (Fig.~S7), the two probes were recorded on the same system using separate hybrid (HyD) detectors: the white-light laser provided 488 nm excitation for FITC-DHPE, whose emission was collected over 493-526 nm (green channel), while the Texas Red-DHPE emission was collected over 601-640 nm (red channel). Co-localization between the two channels was quantified from the raw images in Fiji (ImageJ) \cite{schindelin2012fiji} using the Coloc2 plugin \cite{bolte2006guided} with Costes automatic intensity thresholding \cite{costes2004automatic}, reporting both the intensity-correlation (Pearson's $R$) and the spatial-overlap (Manders' $M_1$, $M_2$) coefficients \cite{manders1993measurement}, with a Costes randomization test to assess statistical significance \cite{costes2004automatic}. This analysis yielded $R = 0.90$, $M_1 \approx 1.0$ and $M_2 \approx 1.0$, with the Costes randomization test confirming non-random spatial correlation ($P = 1.00$). These measurements show that the two spectrally distinct DHPE probes reproduce the same interfacial stripe pattern (Fig.~S7). We note that they do not distinguish the local distributions of DLPC and DOPC.

\subsection{Fluorescence Recovery After Photobleaching}

\indent \indent To investigate the lipid mobility and dynamic properties of the lipid systems doped with 2.8 wt.-\% CB15, FRAP experiments were performed using a Nikon Eclipse Ti inverted microscope. FRAP was carried out on this separate widefield system rather than the confocal used for structural imaging, because the measurement requires fast frame rates, high detector sensitivity, and a targeted-bleaching capability; consequently, the images in Fig.~5 have lower spatial resolution and a different field of view than the confocal micrographs in Figs.~1-4, reflecting the imaging instrument rather than any change in the sample. Fluorescence signals were captured using an Andor iXon DU-897 electron-multiplying charge-coupled device (EMCCD) camera (512 $\times$ 512, 16 µm pixels), which provides the high sensitivity and frame rates required for accurate dynamic tracking. Imaging and targeted photobleaching were conducted through a 40$\times$ oil-immersion objective.

Time-lapse image series were acquired at a fast scan rate with a precise time interval of 0.5 s per frame. To ensure reliable quantitative analysis, all images were recorded at a resolution of 512 $\times$ 512 pixels with a 16-bit depth. Photobleaching of the target region of interest (ROI) was carried out over 20 iterations. Following the bleach pulse, fluorescence recovery was continuously monitored for 205 frames ($\sim102.5$~s). Because recovery did not reach a clear plateau within this interval, the endpoint of the recording was not treated as an asymptotic recovery level.

Post-acquisition, the time-series stacks were concatenated and exported as .tif files using ImageJ/Fiji. Two complementary analyses were then performed. The representative normalized recovery curve shown in Fig.~\ref{fig:s8} was obtained using simFRAP in ImageJ. For this analysis, the characteristic recovery time $\tau_i$ was related to the effective pixel size $l$ through $D=l^2/(4\tau_i)$. Because the fluorescence recovery did not reach a plateau during the measurement window, an immobile fraction was not determined. The diffusion coefficients reported in Fig.~5C were obtained independently using a custom MATLAB (MathWorks) implementation adapted from the method of J\"{o}nsson et al. \cite{jonsson2008method}. This approach analyzes the spatial evolution of the bleached fluorescence profile using a Hankel-transform-based treatment rather than extracting the diffusion coefficient from a fit to the mean recovery curve. Background and acquisition-induced photobleaching corrections were applied using background and unbleached reference regions before quantitative analysis.

\newpage
\section*{Acknowledgements}

\indent \indent M.L. acknowledges support from the China Scholarship Council. M.L. acknowledges the use of Anthropic's Claude for assistance with the Python and LaTeX scripts used for figure preparation and typesetting, and for drafting the text. All experimental work, parameters, and reported values were determined and verified by the authors. All authors have reviewed and edited the final work and take full responsibility for the content. S.D.P. acknowledges support from the European Research Council through the ERC Consolidator Grant ProForce awarded to Jan Lipfert (Grant No. 101002656) and from the Alexander von Humboldt Foundation through a Feodor Lynen Fellowship. S.D.P. acknowledges Jan Lipfert and Alptu\u{g} Ulug\"ol for fruitful discussions. Jan Lipfert is also acknowledged for access to the microscope and EMCCD camera used for FRAP measurements. L.T. acknowledges support from the European Commission (Horizon-MSCA, Grant No. 892354), the Dutch Research Council NWO ENW Veni grant (Project No. VI.Veni.212.028), and the Dutch Research Council NWO ENW M grant (Project No. OCENW.M.23.360).

\section*{Author Contributions}
\indent \indent M.L.: conceptualization, investigation, methodology, formal analysis, visualization, writing--original draft. S.D.P.: investigation and formal analysis of the Fluorescence Recovery After Photobleaching (FRAP) experiments, writing--review and editing. M.F.H.: supervision, resources, writing--review and editing. L.T.: conceptualization, supervision, funding acquisition, writing--review and editing. All authors have given approval to the final version of the manuscript.

\section*{Notes}
\indent \indent The authors declare no competing financial interest. 

\section*{Data Availability}

The data supporting this article are provided in the Supporting Information (SI) and in the associated data repository. The SI contains the molecular structures and sample geometry, additional confocal fluorescence image series and intensity profiles, the two-color fluorescence control, a representative FRAP recovery analysis, and DLS characterization of vesicle stability. The associated data repository contains the confocal image series, two-color fluorescence data, FRAP time series and extracted diffusion coefficients, DLS measurements, and stripe-spacing measurements used in the analysis. The data repository is available through DataverseNL at DOI: \url{https://doi.org/10.34894/KQ9TT4}.

\clearpage
\section*{Supporting Information}

\setcounter{figure}{0} 
\renewcommand{\thefigure}{S\arabic{figure}} 

\indent \indent The Supporting Information includes the molecular structures of the lipids, fluorescent probes, and liquid-crystal components together with the TEM-grid sample geometry (Fig.~S1); a cross-sectional schematic mapping the bright and dark interfacial fingerprint stripes onto the underlying surface anchoring geometry, with the cholesteric helix axis parallel to the interface (Fig.~S2); Z-projected confocal micrographs, with the corresponding grayscale intensity profiles, of the cholesteric interfacial textures as a function of DLPC concentration (Fig.~S3) and of DOPC concentration (Fig.~S4), and of CB15 concentration at fixed DLPC (Fig.~S5) and at fixed DOPC (Fig.~S6); a two-color control of an equimolar (1:1) mixed DLPC:DOPC interface using the spectrally distinct DHPE-based probes Texas Red-DHPE and FITC-DHPE, showing that both fluorescence channels reproduce the same interfacial stripe pattern (Fig.~S7); a representative normalized fluorescence recovery curve with the simFRAP analysis used to illustrate it (Fig.~S8); and dynamic light scattering (DLS) characterization of DLPC vesicle stability during cold storage (Fig.~S9).

\subsection*{Origin of fluorescence contrast in liquid-crystal-side imaging}

\indent \indent The confocal micrographs in Figs.~1-4 were acquired with excitation and fluorescence collection through the cholesteric liquid crystal. This configuration produces two forms of contrast that require careful interpretation. At 1~µM lipid concentration, broad schlieren-like patterns appear in Figs.~1A-i and 1B-i but are barely visible during water-side imaging. At higher concentrations, the narrow twist-disclination lines separating broadened homeotropic regions appear bright through the cholesteric but dark through the aqueous phase. Fluorescence intensity measured through the cholesteric is therefore influenced by the director configuration and cannot be interpreted directly as a quantitative map of lipid concentration.

Several observations show that the bright twist-disclination lines do not represent lipid enrichment. In our previous study of the same lipid-cholesteric interface, water-side confocal imaging showed dark twist-disclination lines relative to the surrounding homeotropic regions \cite{tran2018shaping}. The same contrast was observed in the present study by wide-field fluorescence microscopy with the sample inverted, so that excitation and collection occurred through the aqueous phase. Because wide-field imaging does not involve confocal pinhole rejection or $z$-stack reconstruction, the contrast inversion cannot be attributed to these procedures. Instead, both water-side configurations indicate that the twist-disclination regions contain less fluorescent lipid than the adjacent homeotropically anchored regions.

A two-color control was used to determine whether the striped fluorescence pattern could arise from the photophysical behavior of a particular fluorescent probe. An equimolar mixed DLPC:DOPC interface was imaged using the spectrally distinct DHPE-based probes Texas Red-DHPE and FITC-DHPE. Both fluorescence channels reproduced the same stripe pattern and showed strong spatial correlation (Fig.~S7; Pearson correlation coefficient $R=0.90$). Because both probes are DHPE derivatives and neither is specific to DLPC or DOPC, this experiment does not report the spatial distributions of the two phosphatidylcholine species separately. Instead, it demonstrates that the stripe morphology is reproduced in a second fluorescence channel and is therefore unlikely to arise solely from the partitioning or photophysical behavior of a single fluorophore. Because both channels are imaged through the same birefringent director field, their absolute intensities should still not be interpreted as direct quantitative measures of local lipid concentration.

The bright twist-disclination lines remain visible for a broad range of in-plane orientations and with both fluorescent labels, making a simple dependence on line orientation relative to a fixed excitation polarization unlikely. Imaging was performed using a $93\times$ glycerol-immersion objective with NA = 1.30. Because this numerical aperture is below the refractive index of the aqueous phase, collection of fluorescence emitted into supercritical angles at an ideal planar interface is also unlikely to explain the contrast inversion.

The most consistent explanation is that the cholesteric director field modifies both the excitation of the interfacial fluorophores and the propagation of their emitted light. Director distortions near twist-disclination lines can alter the polarization, phase, wavefront, and focal intensity of the excitation light before it reaches the fluorescent interface. Because fluorescence excitation depends on the orientation of the optical electric field relative to the fluorophore absorption transition dipole, these propagation-induced changes can modify the excitation efficiency of the interfacial probes. The emitted fluorescence is likewise orientation-dependent and can undergo further changes in polarization, angular distribution, transmission, and collection efficiency as it propagates back through the cholesteric. The observed intensity therefore reflects a coupling between the local cholesteric director configuration, fluorophore transition-dipole orientation, and the optical path through the birefringent liquid crystal. This type of orientation-dependent fluorescence response underlies fluorescence confocal polarizing microscopy, in which liquid crystal director configurations are visualized through polarization-sensitive fluorescence intensity \cite{smalyukh2001three,lavrentovich2003fluorescence}, while refractive-index heterogeneity and optical aberrations can further modify excitation and collection efficiency \cite{booth1998aberration,booth2001refractive}. A lipid-depleted twist-disclination region can therefore appear brighter than its surroundings if its local director configuration increases the excitation or detected emission efficiency. Although the present measurements do not uniquely determine the fluorophore transition-dipole orientation because excitation polarization and fluorescence collection were not systematically varied, the strong contrast demonstrates that propagation through the cholesteric provides an additional orientation-sensitive optical readout of the interfacial structure.

The broad schlieren-like patterns at 1~µM are likely produced by the same propagation-dependent mechanism. At this low surface coverage, lipid adsorption weakly perturbs the predominantly planar interface but does not yet establish a well-developed fingerprint texture. Smooth variations in director orientation and pitch-axis direction can consequently produce broad intensity variations when viewed through the birefringent cholesteric. Their weak appearance during water-side imaging indicates that they primarily reveal director distortions rather than large-scale variations in lipid concentration. Thus, liquid-crystal-side imaging appears to provide optical sensitivity to the onset of interfacial director reorganization even before distinct fingerprint stripes emerge.

The water-side wide-field configuration was also used for the FRAP measurements in Fig.~5. The sample was inverted and imaged on a separate microscope, with excitation and collection through the aqueous phase. Under these conditions, twist-disclination lines appeared dark, consistent with lower fluorescent lipid abundance in these regions. The FRAP measurements were therefore not subject to the contrast inversion observed during liquid-crystal-side confocal imaging and directly report the lateral redistribution of the fluorescent probe within the interfacial lipid layer.

Together, water-side confocal imaging, water-side wide-field microscopy, two-color co-localization, and liquid-crystal-side confocal imaging establish that the observed contrast contains both compositional and director-dependent optical contributions. The dark water-side twist-disclination lines indicate lower fluorescent lipid abundance, whereas their bright appearance through the cholesteric arises predominantly from propagation through the anisotropic director field. The schlieren-like contrast at low concentration similarly provides evidence of weak director distortions preceding the development of a fingerprint texture. Accordingly, liquid-crystal-side fluorescence imaging can reveal the morphology and evolution of the cholesteric director field, while water-side measurements provide the more direct indication of lipid localization and mobility. These complementary configurations support the interpretation of the observed transitions as consequences of competition between lipid-induced homeotropic anchoring and bulk cholesteric twist elasticity.

\begin{figure}[H]
 \centering
    \captionsetup{font=scriptsize}
    \centering
    \includegraphics[width=0.9\linewidth]{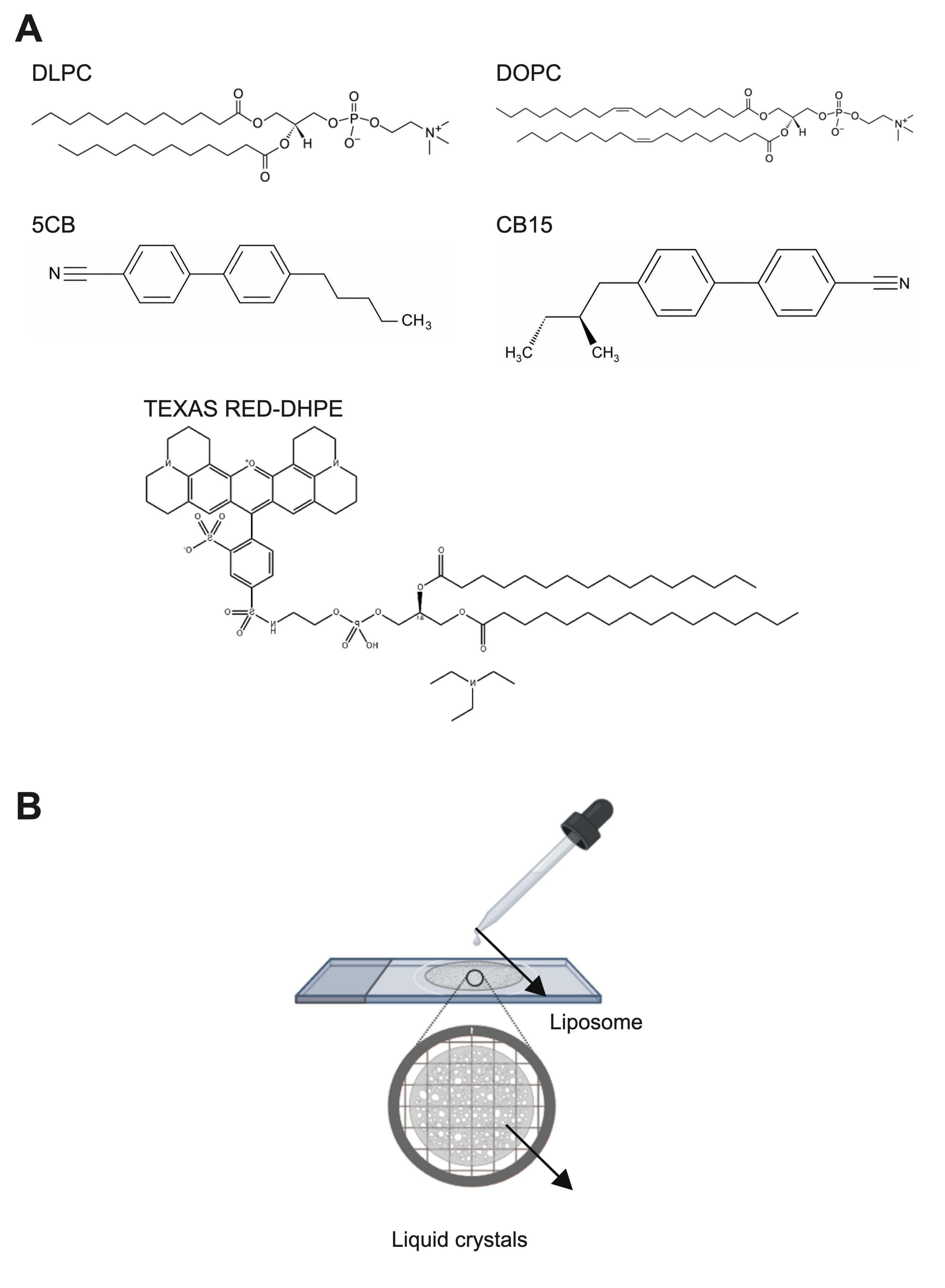}
    \caption{Materials and experimental geometry. (A) Molecular structures of the phospholipids (DLPC, DOPC), the fluorescent probe (Texas Red-DHPE), and the liquid-crystal components (5CB host and the chiral dopant CB15). (B) Sample geometry: the cholesteric liquid crystal is confined by a copper TEM grid resting on an OTS-treated glass slide, and lipid vesicles introduced into the surrounding aqueous well deposit lipids at the liquid crystal-water interface upon rupture.}
    \label{fig:s1}
\end{figure}

\begin{figure}[H]
 \centering
    \captionsetup{font=scriptsize}
    \centering
    \includegraphics[width=0.92\linewidth]{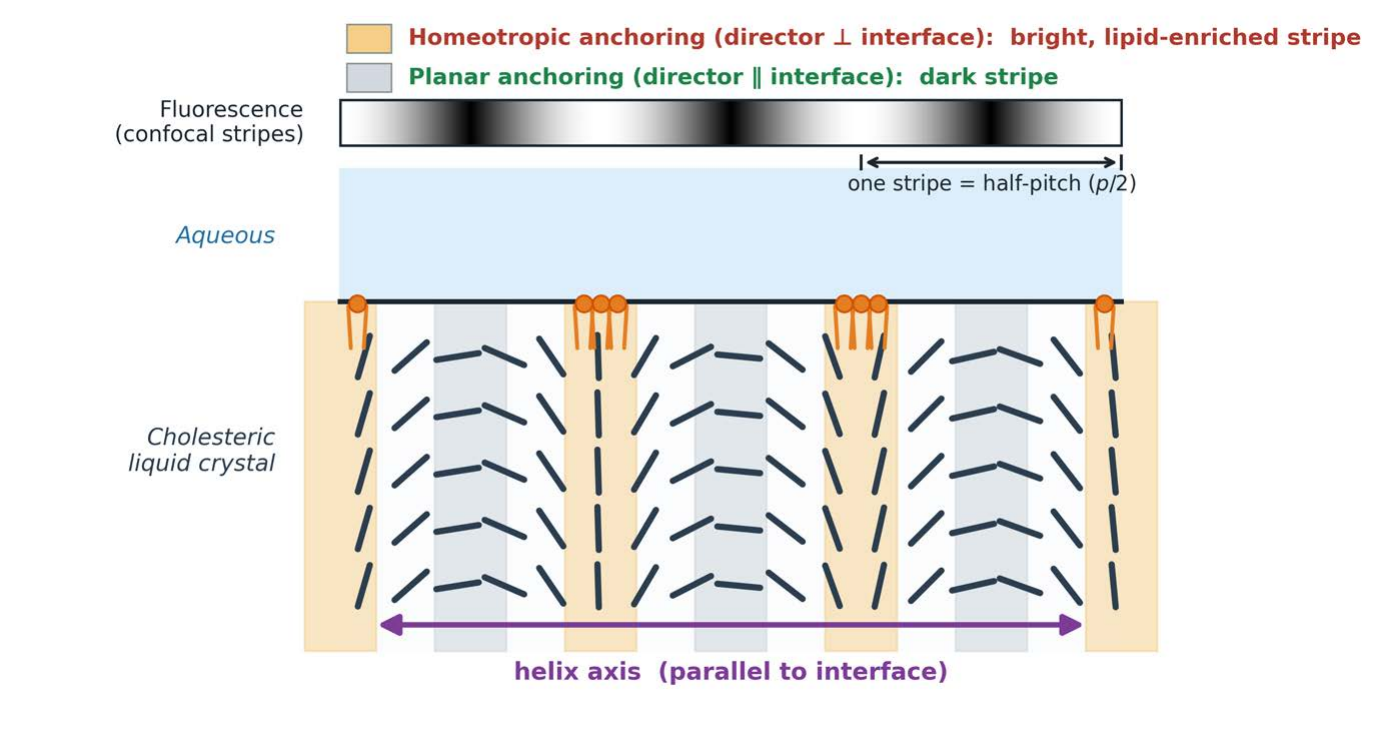}
    \caption{Schematic (side-view cross-section) mapping the interfacial fingerprint stripes onto the underlying surface anchoring geometry. The cholesteric helix axis lies parallel to the aqueous interface, so the director rotates along the interface and the surface anchoring alternates between homeotropic (director normal to the interface), where the lipid tails intercalate into the liquid crystal and the fluorescence is bright (lipid-enriched, amber columns), and planar (director in the interface plane), where the fluorescence is dark (grey columns). The director is drawn as rods that rotate in the plane of the page, vertical at homeotropic columns and flat at planar columns; the greyscale bar at the top shows the resulting confocal stripe contrast. One bright-to-bright stripe corresponds to a half-pitch ($p/2$) rotation of the helix, so the measured stripe spacing reports the local half-pitch.}
    \label{fig:s2}
\end{figure}

\begin{figure}[H]
 \centering

    \captionsetup{font=scriptsize}
\centering
\includegraphics[width=1\linewidth]{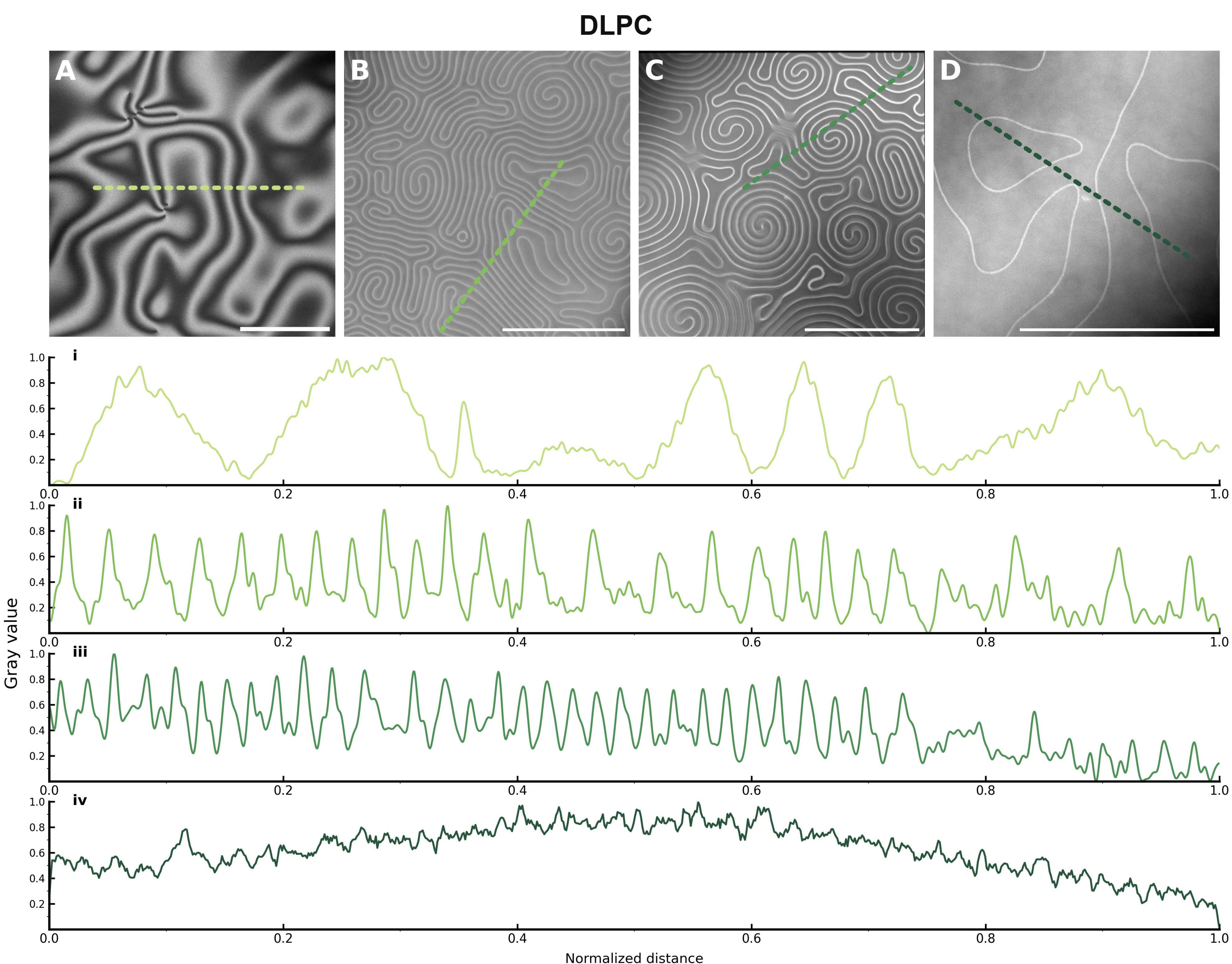}
\caption{Z-projected confocal micrographs of cholesteric liquid crystal films composed of 5CB doped with 2.8 wt.-\% CB15 in the presence of DLPC at different concentrations: (A) 1 µM, (B) 10 µM, (C) 30 µM, and (D) 50 µM. The dashed lines in each image indicate the positions used to obtain the corresponding grayscale intensity profiles shown in (i-iv). These profiles illustrate the concentration-dependent changes in stripe periodicity and contrast across the interface.}
\label{fig:s3}
\end{figure}

\begin{figure}[H]
 \centering
    
    \captionsetup{font=scriptsize}
\centering
\includegraphics[width=1\linewidth]{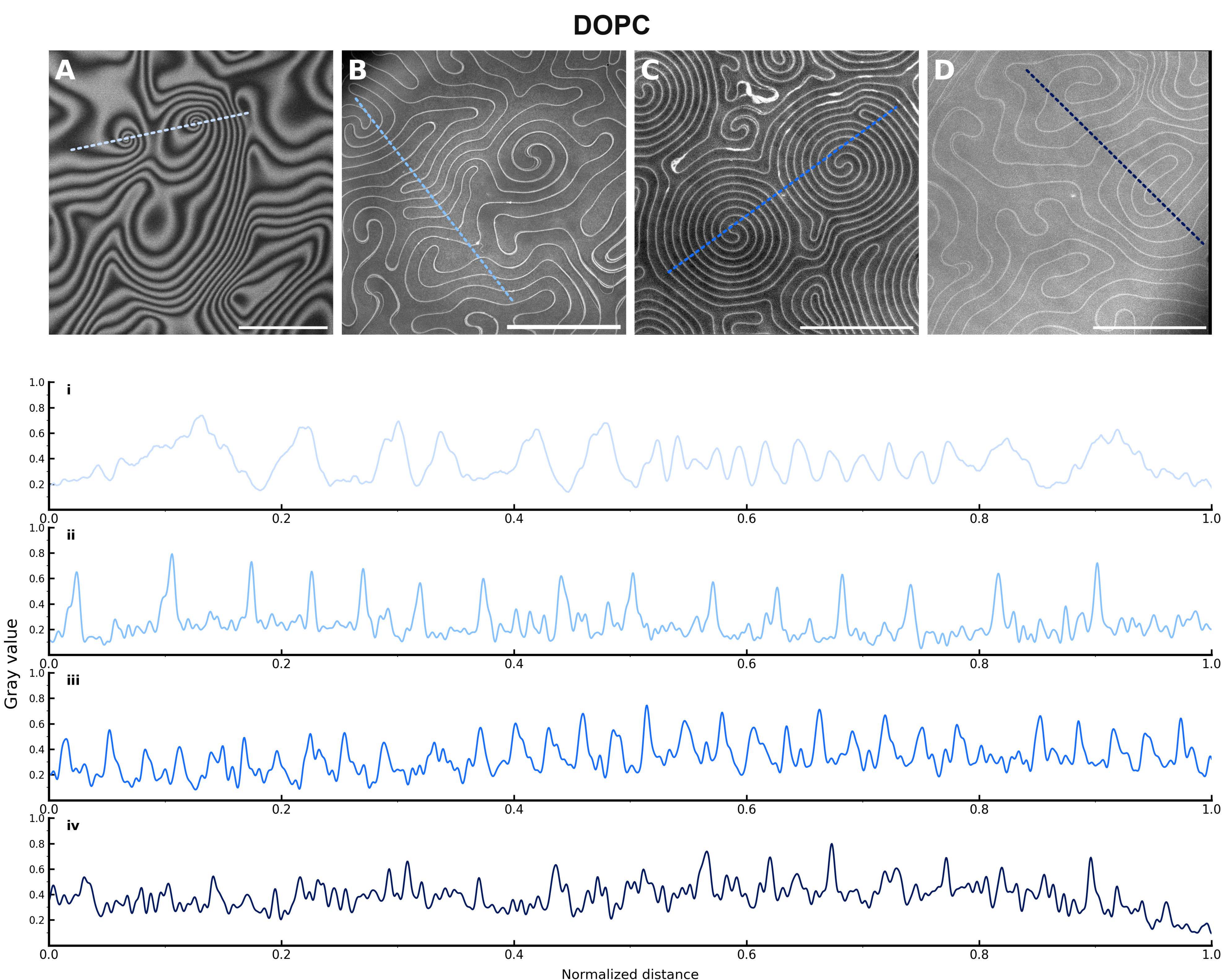}
\caption{Z-projected confocal micrographs of cholesteric liquid crystal films composed of 5CB doped with 2.8 wt.-\% CB15 in the presence of DOPC at different concentrations: (A) 1 µM, (B) 10 µM, (C) 30 µM, and (D) 50 µM. The dashed lines in panels A-D indicate the positions used to extract the corresponding grayscale intensity profiles shown in (i-iv), respectively. The images and intensity traces show concentration-dependent changes in the interfacial stripe pattern.}
\label{fig:s4}
\end{figure}

At a fixed chiral dopant concentration of 2.8 wt.-\%, corresponding to a long pitch of approximately 5 µm, the stripe patterns are visible across 1, 10, 30, and 50 µM DOPC, while the corresponding grayscale intensity profiles in i-iv reflect the stripe modulation extracted along the dashed lines in panels A-D.

\begin{figure}[H]
 \centering
    
    \captionsetup{font=scriptsize}
    \centering
    \includegraphics[width=1\linewidth]{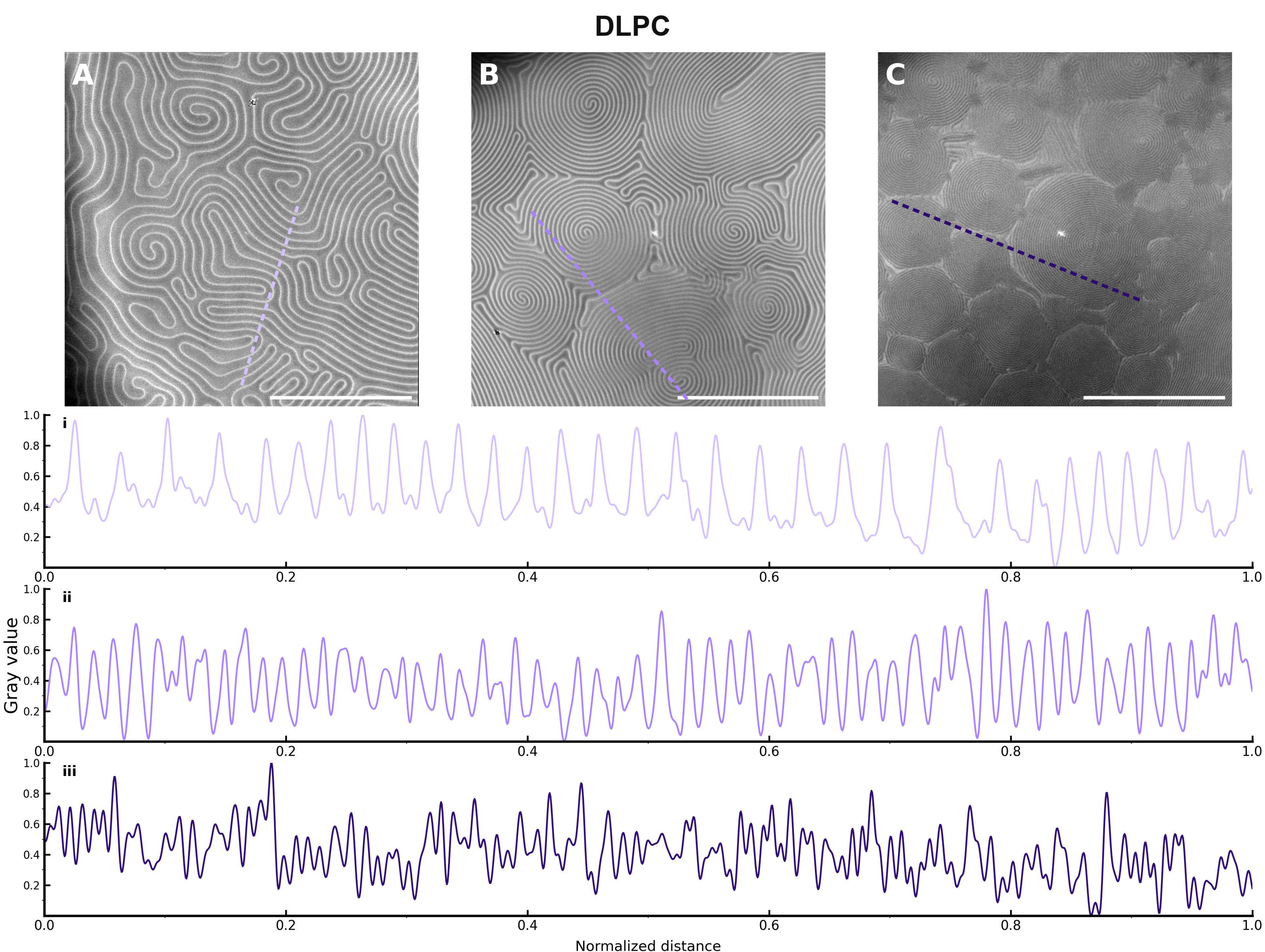}
    \caption{Z-projected confocal micrographs of cholesteric liquid crystal films in the presence of 10 µM DLPC with different CB15 concentrations: (A) 5CB doped with 2.8 wt.-\% CB15, (B) 5CB doped with 5 wt.-\% CB15, and (C) 5CB doped with 10 wt.-\% CB15. The dashed lines in panels A-C indicate the positions used to extract the corresponding grayscale intensity profiles shown in (i-iii), respectively. The intensity traces show a decrease in stripe spacing with increasing CB15 concentration.}
    \label{fig:s5}
\end{figure}

At a fixed DLPC concentration of 10 µM, increasing the CB15 concentration led to clear changes in the stripe patterns of the cholesteric films. For 5CB doped with 2.8 wt.-\% CB15, the stripe features were relatively broad and well resolved. At 5 wt.-\% CB15, the stripe spacing decreased and the pattern became denser. When the CB15 concentration was further increased to 10 wt.-\%, the characteristic features became more closely packed and less continuous. The grayscale intensity profiles extracted along the dashed lines in panels A-C, shown in (i-iii), respectively, further confirm the reduction in stripe spacing with increasing chiral dopant content.

\begin{figure}[H]
 \centering
    
    \captionsetup{font=scriptsize}
    \centering
    \includegraphics[width=1\linewidth]{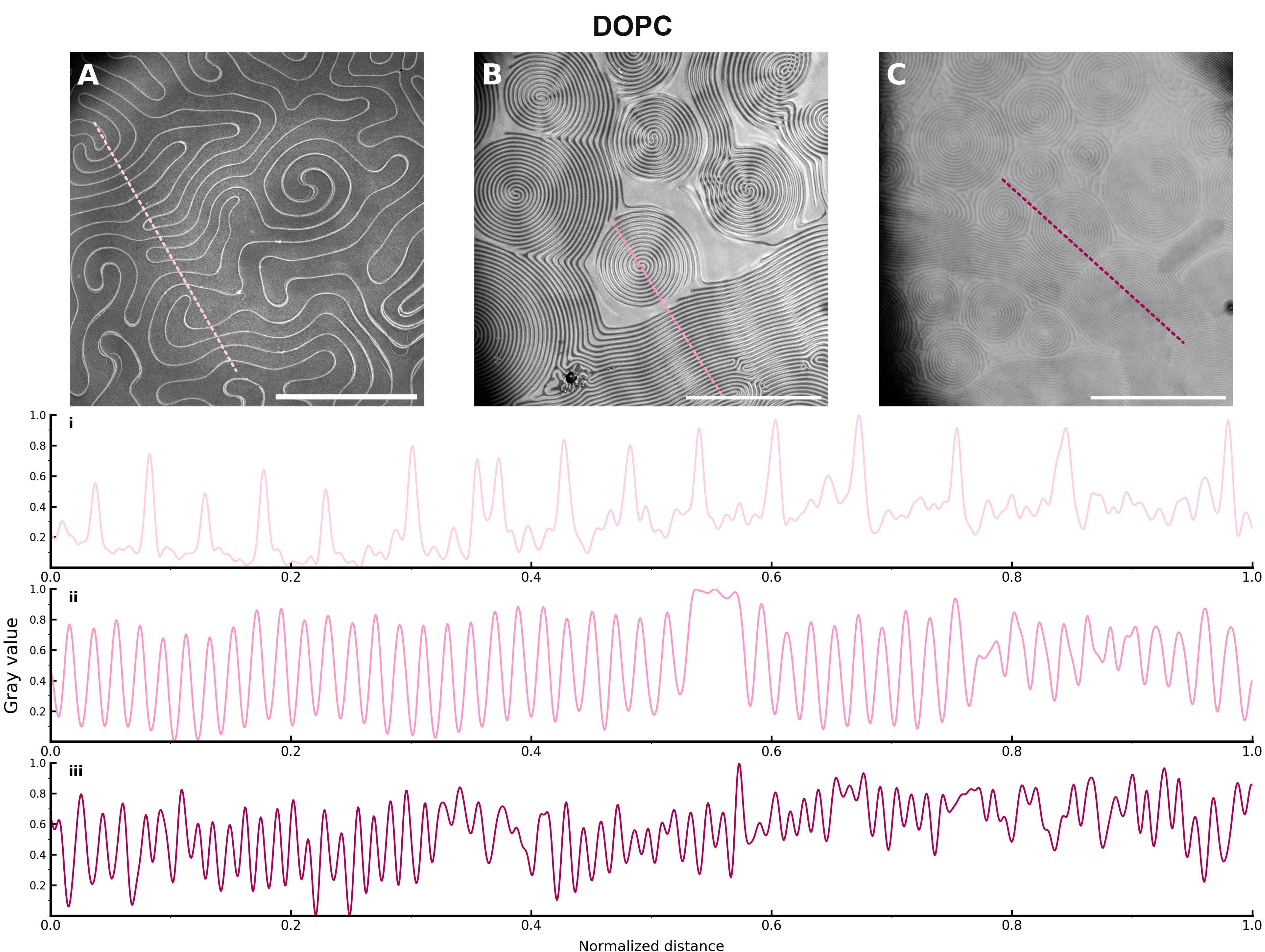}
    \caption{Z-projected confocal micrographs of cholesteric liquid crystal films in the presence of 10 µM DOPC with different CB15 concentrations: (A) 5CB doped with 2.8 wt.-\%  CB15, (B) 5CB doped with 5 wt.-\%  CB15, and (C) 5CB doped with 10 wt.-\%  CB15. The dashed lines in panels A-C indicate the positions used to extract the corresponding grayscale intensity profiles shown in (i-iii), respectively. The stripe spacing decreases progressively with increasing CB15 concentration.}
    \label{fig:s6}
\end{figure}
At a fixed DOPC concentration of 10 µM, increasing the CB15 concentration leads to a progressive decrease in stripe spacing. At 2.8 wt.-\% CB15, the stripe patterns are widely spaced and remain clearly visible at the interface. When the CB15 concentration is increased to 5 wt.-\%, the stripe spacing becomes smaller and the fingerprint texture becomes denser. At 10 wt.-\% CB15, the most closely spaced stripe pattern is observed. The corresponding grayscale intensity profiles extracted along the dashed lines in panels A-C, shown in i-iii, respectively, reflect this progressive reduction in stripe spacing with increasing chiral dopant concentration.

\begin{figure}[H]
 \centering

    \captionsetup{font=scriptsize}
    \centering
    \includegraphics[width=0.8\linewidth]{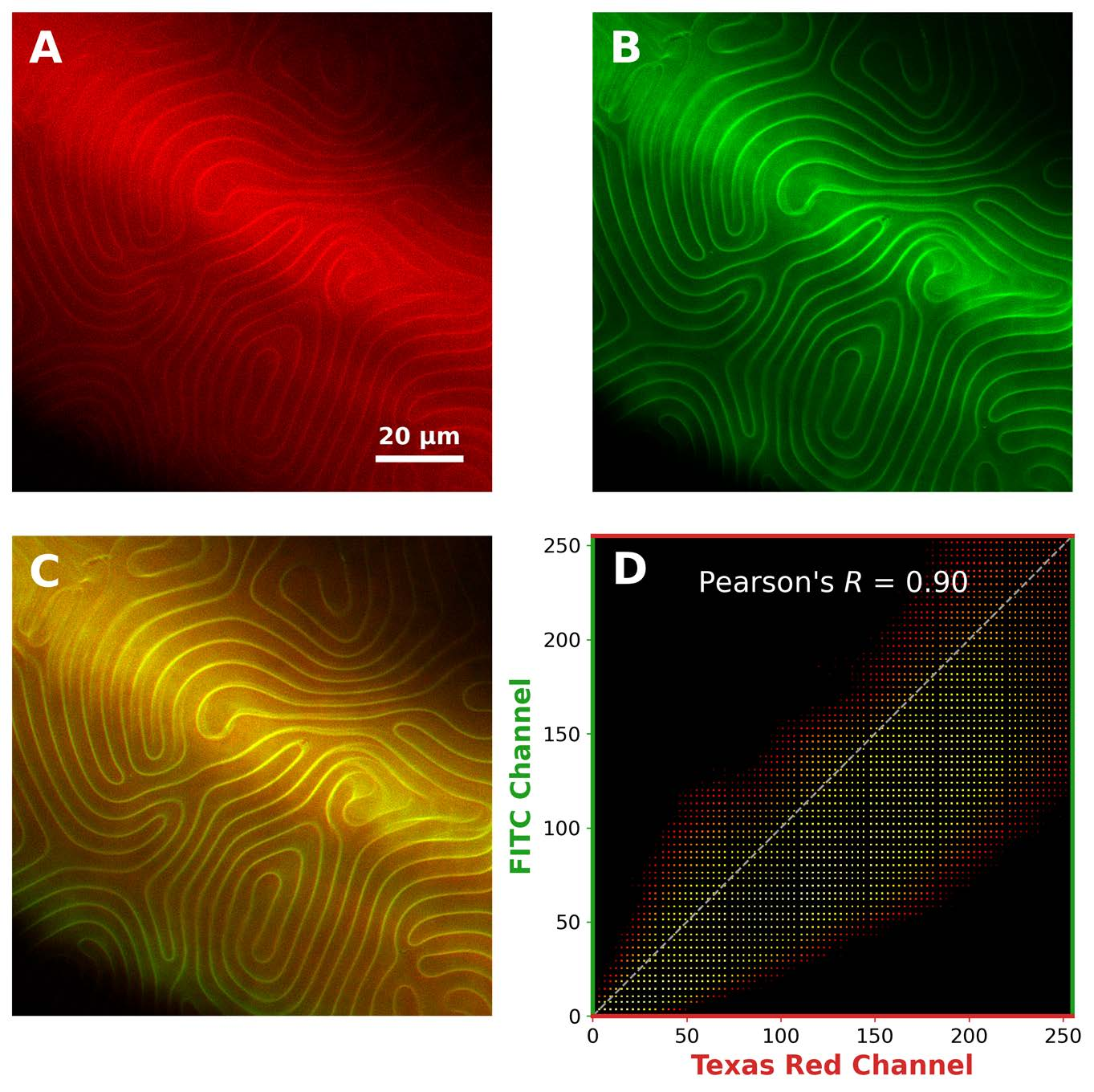}
    \caption{Two-color fluorescence control of an equimolar (1:1) mixed DLPC:DOPC interface at a total lipid concentration of 10~µM using two spectrally distinct DHPE-based fluorescent probes. (A) Texas Red-DHPE channel. (B) FITC-DHPE channel. (C) Merge of the two channels. (D) Pixel-by-pixel intensity scatter plot of Texas Red-DHPE versus FITC-DHPE fluorescence; dashed line indicates identity. Each single-channel image is a deconvolved confocal average projection. Scale bar: 20~µm.}
    \label{fig:s7}
\end{figure}

Figure S7. The two probes were excited and detected separately (FITC-DHPE, 488 nm excitation, 493-526 nm emission; Texas Red-DHPE, 601-640 nm emission), and co-localization was quantified from the raw, unstretched images in Fiji (ImageJ) using the Coloc2 plugin with Costes automatic intensity thresholding \cite{schindelin2012fiji, bolte2006guided, costes2004automatic, manders1993measurement}. The two channels are strongly correlated, with a Pearson correlation coefficient of $R = 0.90$ (the scatter density in panel D accordingly lies along the diagonal), and essentially all of each probe's signal overlaps the other, with Manders' overlap coefficients $M_1 \approx 1.0$ and $M_2 \approx 1.0$. A Costes randomization test confirmed that this correlation is statistically significant rather than a chance overlap ($P = 1.00$; i.e., the measured $R$ exceeds that of all randomized-image controls). Because Pearson's $R$ reports co-variation of the two fluorescence intensities while Manders' coefficients report their spatial overlap, these measurements demonstrate that Texas Red-DHPE and FITC-DHPE reproduce the same interfacial stripe pattern. They do not establish the local distributions of DLPC and DOPC because neither probe is chemically specific to either lipid species.

\begin{figure}[H]
    \centering
    \captionsetup{font=scriptsize}
    \includegraphics[width=0.62\linewidth]{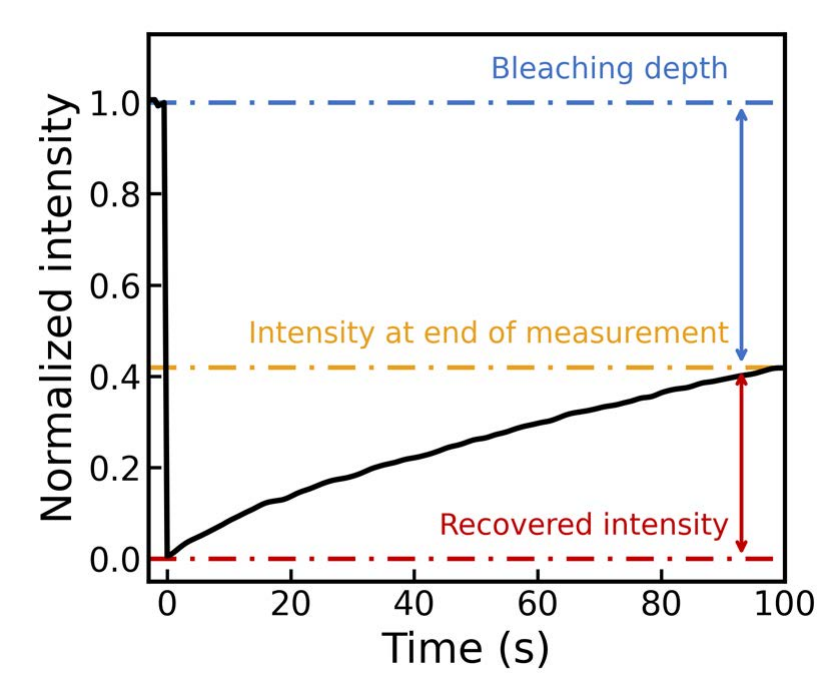}
    \caption{Representative normalized fluorescence recovery curve for the measurement shown in Fig.~5A, obtained with simFRAP in ImageJ. The pre-bleach level, the post-bleach baseline, and the intensity reached at the end of the recording are indicated. The recovery does not reach a plateau within the approximately 100~s measurement window, so the terminal intensity represents the level attained at the end of the recording rather than an asymptotic recovery level, and no immobile fraction is reported. Within this pixel-based analysis a characteristic recovery time $\tau_i$ can be related to a diffusion coefficient by $D = l^2/(4\tau_i)$, where $l = 0.4$~µm is the effective pixel size for the 40$\times$ imaging configuration; this relation follows the two-dimensional mean-squared-displacement framework used in pixel-grid FRAP analyses \cite{blumenthal2015universal} and is consistent with the general relationship between recovery time and lateral diffusion in classical FRAP treatments \cite{axelrod1976mobility,soumpasis1983theoretical}. The diffusion coefficients reported in Fig.~5C were obtained independently with the J\"{o}nsson analysis and do not derive from this curve.}
    \label{fig:s8}
\end{figure}

\begin{figure}[H]

 \centering

    \captionsetup{font=scriptsize}
\centering
\includegraphics[width=1\linewidth]{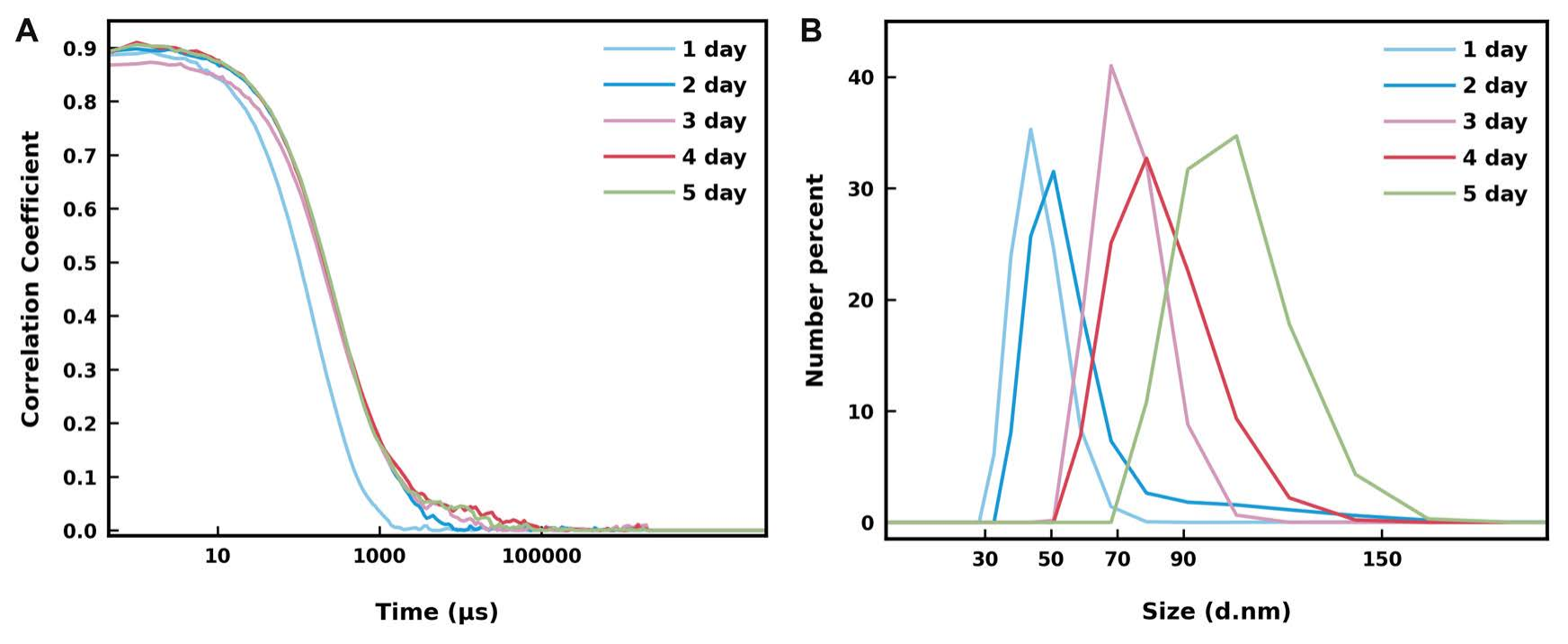}
\caption{Dynamic light scattering (DLS) analysis of DLPC vesicles stored at 4 °C for 5 days. (A) Autocorrelation curves of DLPC vesicles measured after different storage times (1-5 days). (B) Corresponding number-based size distributions as a function of hydrodynamic diameter (d.nm).}
\label{fig:s9}
\end{figure}

Figure S9. Dynamic light scattering (DLS) characterization of DLPC vesicles stored at 4 °C for up to 5 days. These results indicate that the vesicle dispersion remained generally stable during cold storage, with moderate time-dependent size growth.

\clearpage

\bibliography{references}

\end{document}